\documentclass[11pt]{article}

\newsavebox{\foobox}
\newcommand{\slantbox}[2][0]{\mbox{%
        \sbox{\foobox}{#2}%
        \hskip\wd\foobox
        \pdfsave
        \pdfsetmatrix{1 0 #1 1}%
        \llap{\usebox{\foobox}}%
        \pdfrestore
}}
\newcommand\unslant[2][-.25]{\slantbox[#1]{$#2$}}

\newcommand{\mpi}{\text{\unslant[-.18]\pi}}
\newcommand{\mdelta}{\text{\unslant[-.18]\delta}}

\usepackage[left=2cm, right=2cm, top=2.5cm, bottom=2.5cm]{geometry}
\usepackage[x11names]{xcolor}
\usepackage{fancyhdr, amssymb, cancel, amsmath, graphicx, pgfplots, tikz}
\usepackage{isomath}

\usetikzlibrary{shadows}

\newcommand{\stylecolor}{violet}

\usepackage[labelfont={bf,sf, color=\stylecolor}, margin={1.5cm,0cm}]{caption}

\usepackage[colorlinks=true, urlcolor=\stylecolor!70!white, linkcolor=\stylecolor, citecolor=\stylecolor!70!white, hyperindex=true, linktocpage=true]{hyperref}

\usepackage[explicit]{titlesec}

\newcommand*\sectionlabel{}
\titleformat{\section}
  {\gdef\sectionlabel{}
   \Large\bfseries\scshape}
  {\gdef\sectionlabel{\thesection }}{0pt}
  {\begin{tikzpicture}[remember picture,overlay]
       \end{tikzpicture}
  }
\titlespacing*{\section}{0pt}{0pt}{0pt}

\newcommand*\subsectionlabel{}
\titleformat{\subsection}
  {\gdef\subsectionlabel{}
   \large\bfseries\scshape}
  {\gdef\subsectionlabel{\thesubsection  }}{0pt}
  {\begin{tikzpicture}[remember picture,overlay]
    	\draw (-0.15, 0.02) node[right] {\color{\stylecolor} \textsf{\subsectionlabel}};
	\draw (1.25, 0) node[right] {\color{\stylecolor} \textsf{#1}};
	\fill[color=\stylecolor] (1,-0.25) rectangle (1.1, 0.25);
       \end{tikzpicture}
  }
\titlespacing*{\subsection}{0pt}{10pt}{10pt}

\newcommand*\subsubsectionlabel{}
\titleformat{\subsubsection}
  {\gdef\subsubsectionlabel{}
   \bfseries\scshape}
  {\gdef\subsubsectionlabel{\thesubsubsection.\ \  }}{0pt}
  {\begin{tikzpicture}[remember picture,overlay]
    	\draw (-0.15, 0) node[right] {\color{\stylecolor} \textsf{\subsubsectionlabel#1}};
       \end{tikzpicture}
  }
\titlespacing*{\subsubsection}{0pt}{7pt}{7pt}

\pgfplotsset{every axis legend/.append style={at={(1.02,1)},anchor=north west}}

\begin{document}

\allowdisplaybreaks

\pagestyle{fancy}
\renewcommand{\headrulewidth}{0pt}
\fancyhead{}

\fancyfoot{}
\fancyfoot[C] {\textsf{\textbf{\thepage}}}

\begin{equation*}
\begin{tikzpicture}
\draw (\textwidth, 0) node[text width = \textwidth, right] {\color{white} easter egg};
\end{tikzpicture}
\end{equation*}

\begin{equation*}
\begin{tikzpicture}
\draw (0.5\textwidth, -3) node[text width = \textwidth] {\huge  \textsf{\textbf{Hydrodynamic transport in strongly coupled  \\  \vspace{0.07in}  disordered quantum field theories}} };
\end{tikzpicture}
\end{equation*}
\begin{equation*}
\begin{tikzpicture}
\draw (0.5\textwidth, 0.1) node[text width=\textwidth] {\large \color{black} $\text{\textsf{Andrew Lucas}}$};
\draw (0.5\textwidth, -0.5) node[text width=\textwidth] {\small \textsf{Department of Physics, Harvard University, Cambridge, MA 02138, USA}};
\end{tikzpicture}
\end{equation*}
\begin{equation*}
\begin{tikzpicture}
\draw (0, -13.1) node[right, text width=0.5\paperwidth] {\texttt{lucas@fas.harvard.edu}};
\draw (\textwidth, -13.1) node[left] {\textsf{\today}};
\end{tikzpicture}
\end{equation*}
\begin{equation*}
\begin{tikzpicture}
\draw[very thick, color=\stylecolor] (0.0\textwidth, -5.75) -- (0.99\textwidth, -5.75);
\draw (0.12\textwidth, -6.25) node[left] {\color{\stylecolor}  \textsf{\textbf{Abstract:}}};
\draw (0.53\textwidth, -6) node[below, text width=0.8\textwidth, text justified] {\small We compute direct current (dc) thermoelectric transport coefficients in strongly coupled quantum field theories without long lived quasiparticles, at finite temperature and charge density, and disordered on long wavelengths compared to the length scale of local thermalization.    Many previous transport computations  in strongly coupled systems are interpretable hydrodynamically, despite formally going beyond the hydrodynamic regime.   This includes momentum relaxation times previously derived by the memory matrix formalism, and non-perturbative holographic results;  in the latter case, this is subject to some important subtleties.   Our formalism may extend some memory matrix computations to higher orders in the perturbative disorder strength, as well as give valuable insight into non-perturbative regimes.   Strongly coupled metals with quantum critical contributions to transport generically transition between coherent and incoherent metals as disorder strength is increased at fixed temperature, analogous to mean field holographic treatments of disorder.    From a  condensed matter perspective, our theory generalizes the resistor network approximation, and associated variational techniques, to strongly interacting systems where momentum is long lived.};
\end{tikzpicture}
\end{equation*}

\tableofcontents

\titleformat{\section}
  {\gdef\sectionlabel{}
   \Large\bfseries\scshape}
  {\gdef\sectionlabel{\thesection }}{0pt}
  {\begin{tikzpicture}[remember picture,overlay]
	\draw (1, 0) node[right] {\color{\stylecolor} \textsf{#1}};
	\fill[color=\stylecolor] (0,-0.35) rectangle (0.7, 0.35);
	\draw (0.35, 0) node {\color{white} \textsf{\sectionlabel}};
       \end{tikzpicture}
  }
\titlespacing*{\section}{0pt}{15pt}{15pt}

\begin{equation*}
\begin{tikzpicture}
\draw[very thick, color=\stylecolor] (0.0\textwidth, -5.75) -- (0.99\textwidth, -5.75);
\end{tikzpicture}
\end{equation*}

\section{Introduction}

One of the most exotic and mysterious systems in condensed matter physics is the strange metal, non-Fermi liquid phase of the high $T_{\mathrm{c}}$ superconductors \cite{taillefer, keimer}.  The transport data in these materials -- including, most famously, the linear in temperature dc electrical resistivity -- defies clear explanation by a theory of long lived quasiparticles  \cite{kasahara}.   Alternatively, the effectively relativistic plasma in graphene may provide an experimental realization of a strongly interacting quantum fluid \cite{muller}.    Finally, recent advances in ultracold atomic gases have paved the way to realizing strongly interacting fluids \cite{adams}.    In all of the above systems, the absence of long lived quasiparticles on experimentally appropriate time scales (e.g., in the computation of dc transport coefficients) poses a challenge for traditional, quasiparticle-based approaches to condensed matter physics.

From a theoretical perspective,  a generic strongly interacting quantum field theory (QFT) in more than one spatial dimension has only a few quantities (energy, charge and momentum) that are  long lived, and so hydrodynamics may be a sensible description of the low energy physics at finite temperature and density of all of the above systems.  Though hydrodynamics is an old theory \cite{landau, kadanoff}, its implications for transport have only been understood comparatively recently \cite{hkms}, in weakly disordered systems near quantum criticality, in external magnetic fields \cite{hkms, bhaseen, bhaseen2}, and  in some simple examples of disordered, non-relativistic electron fluids \cite{andreev}.    This is because ``textbook" hydrodynamics is utterly inappropriate for most metals,  where the electron-impurity scattering length is short compared to the electron-electron scattering length.   Momentum and energy rapidly decay, and the only hydrodynamic variable is the charge density.   Note that in contrast to this canonical lore, \cite{zaanen} proposed that observing viscous hydrodynamics in some metals may be possible.   

In most ways, hydrodynamics is a far simpler theory to understand (and perform computations in) than quasiparticle based approaches, such as kinetic theory.   The difficulty in studying these systems theoretically lies in the fact that hydrodynamics does not completely solve the transport problem: the coefficients in the hydrodynamic equations must be related to Green's functions in a microscopic model.  Nonetheless, if hydrodynamics is valid, it does provide universal constraints on transport, and a transparent physical picture to interpret the results.   There are two tractable approaches that can compute the requisite microscopic Green's functions, without reference to quasiparticles.  The first is methods from (perturbative) QFT,  combined with the memory matrix approach \cite{zwanzig, mori, forster}, which has recently been used in many microscopic models of strange metals, reasonable for describing cuprate strange metals \cite{raghu1, raghu2, patel, debanjan}.   These approaches rely on properly resumming certain families of Feynman diagrams to all orders.   The second approach is holography \cite{review1, review2, review3}, which reduces the computation of Green's functions to solving classical differential equations in a  black hole background.   This can be done numerically \cite{santoslat1, santoslat2, santoslat3, chesler, ling, donos1409, rangamani}, though in some cases analytic insight can be obtained \cite{ugajin, chesler, btv, lss, donos1409, lucas1411, peet, rangamani, grozdanov}, sometimes by employing the memory matrix method \cite{hkms, hartnollimpure, hartnollhofman}.     

Surprisingly, many of the above transport theories from recent years completely match hydrodynamic predictions, at least superficially, despite being formally beyond the regime of validity of  hydrodynamics.   We  take this as an indication that a thorough understanding of  hydrodynamic implications for transport in disordered theories is worthwhile, though we will also carefully describe the regime of validity of the approach.   In addition, while almost every citation above aims to address the strange metal phase \cite{taillefer, keimer, kasahara},  the ``hydrodynamic insight" gained from these methods may be applicable to a much broader set of experimentally realized  interacting quantum systems.

\subsection{Motivation: Incoherent Metals and Holography}
Let us begin with the main quantitative motivation for the present paper, which is the physical interpretation of a large body of recent holographic work on transport in QFTs without translational symmetry.

\cite{hkms} proposed a simple hydrodynamic framework for dc transport which has been quite predictive of both holographic and memory function results in subsequent works, at weak disorder.     As we previously mentioned,  this framework has been surprisingly good at describing many holographic models that treat disorder at a mean field level.   A natural conjecture is that disordered hydrodynamic dc transport can describe holographic systems with explicitly broken translational symmetry, and so it is worthwhile to fully flesh out the disordered hydrodynamic formalism.

We begin with a generic hydrodynamic framework for zero frequency transport calculations in Section \ref{sec2}.   Our emphasis is on a clear presentation of the assumptions and regime of validity of a hydrodynamic description of transport.   

We exactly solve the transport problem in Section \ref{sec3}, at leading order in the strength of disorder, in the limit where translational symmetry is only broken weakly.    These systems describe coherent metals, in the language of \cite{hartnoll1} -- henceforth we will also adopt this terminology.    We show that our resulting computations exactly equal the results found by the memory function formalism, under the assumption that the momentum is long-lived, justifying that our approach is sensible, as well as providing a physically transparent derivation of memory function based formulas for conductivities (at least, in the mutual regime of validity of the two methods).    

We further show in Section \ref{sec4} that the hydrodynamic framework can be used to interpret exact, non-perturbative analytic results for dc transport found using holography.  This is subject to the important subtlety that the transport coefficients are computed in terms of a new emergent horizon fluid with distinct, emergent (but somewhat sensible) equations of state.  

We then proceed to study hydrodynamic transport in non-perturbatively disordered QFTs.   Though not amenable to analytic techniques, we develop a combination of rigorous variational approaches and heuristic approximations, outlined in Section \ref{sec5}, to calculate dc transport in this regime.  One might expect that transport becomes dominated by dissipative hydrodynamics, as momentum may become a ``bad" conserved quantity;  such a state is an incoherent metal, in the language of \cite{hartnoll1}.  We find further evidence for this qualitative picture.
  

To date, all models of incoherent metals are holographic massive gravity models \cite{vegh, davison, blake1, dsz, thermoel1, thermoel2}, or similarly inspired holographic approaches \cite{hartnolldonos, donos1, andrade, donos2, gouteraux, thermoel3, donos1406, kim, gouteraux2, davison15, blake2} which we will henceforth lump together under the label of mean-field disorder.   These models break translational symmetry phenomenologically, but not explicitly.\footnote{This is technically not quite right -- there is one (set of) bulk scalar fields in these models which is of the form $\phi_i = kx_i$, but this choice maintains homogeneity in the sectors of the theory of interest.}    These models always predict dc transport which, at all disorder strengths, can be interpreted in terms of the hydrodynamic results of \cite{hkms}, or the slight generalization of \cite{lucasMM}.   A simple example of this is the exact formula for dc electrical conductivity in an isotropic system \cite{blake1}: \begin{equation}
\sigma = \sigma_{\textsc{q}} + \frac{\mathcal{Q}^2 \tau}{\mathcal{M}},  \label{drude}
\end{equation}
with $\sigma_{\textsc{q}}$ a transport coefficient independent of disorder strength, $\mathcal{Q}$ a charge density, and $\mathcal{M}$ an analogue of the mass density.   The parameter $\tau$ is analogous to a momentum relaxation time, and related to the phenomenological graviton mass.  This formula was already known from quantum critical hydrodynamics \cite{hkms}, using computations valid as $\tau \rightarrow \infty$.  Indeed, the latter term is nothing more than the Drude formula, valid in a system without quasiparticles, and the former is a quantum effect that can be important close to a particle-hole symmetric point \cite{patel2}.   Mean field models always predict that (\ref{drude}) holds even as $\tau\rightarrow 0$, or in the non-perturbative, strong disorder regime.   In this limit $\tau$ cannot be interpreted as the momentum relaxation time directly, but importantly, $\sigma$ stays larger than $\sigma_{\textsc{q}}$, which is the conductivity when $\mathcal{Q}=0$ (an uncharged theory).\footnote{See \cite{davison15, blake2} for recent updates on this particular holographic model.}   And while mean field models do agree with approaches that explicitly break translational symmetry weakly \cite{btv, lss, lucas1411},  this is simply a consequence of the perturbative equivalence between holographic and memory function computations of transport, proven in many cases in \cite{lucas}.

One might suspect that the fact that (\ref{drude}) holds as $\tau\rightarrow 0$ is a sign that mean field  physics is a poor description of a strongly disordered QFT, even in holography.   For example, it is well known that mean field descriptions of disorder can completely fail to capture even basic thermodynamics of strongly disordered spin models in classical statistical physics -- instead, the emergent phases are spin glasses and must be treated using much more delicate technologies \cite{spinglass}.   

Our work in Section \ref{sec5} demonstrates that our hydrodynamic framework gives an independent framework in which the qualitative picture of dc transport given in  (\ref{drude})   \emph{is correct at all disorder strengths} until the hydrodynamic description fails.   As an important example, we argue that for an isotropic quantum critical system where viscous transport may be neglected, \begin{equation}
\sigma_{\textsc{q}1}(u) + \sigma_{\textsc{q}2}(u)\frac{\mathcal{Q}_0^2}{u^2} \le \sigma \lesssim \sigma_{\textsc{q}3}(u) + \sigma_{\textsc{q}4}(u) \frac{\mathcal{Q}_0^2}{u^2}, \label{eq2}
\end{equation} with $\sigma$ the dc electrical conductivity, $\mathcal{Q}_0$ the spatial average of the charge density $\mathcal{Q}$, $u$ the typical size of fluctuations in $\mathcal{Q}$ about this average, and $\sigma_{\textsc{q}1,2,3,4}$ are related to the ``quantum critical" diffusive conductivity, $\sigma_{\textsc{q}}$.\footnote{In graphene, for example, $\sigma_{\textsc{q}} \sim e^2/h$, with $e$ the charge of the electron.}   As $u\rightarrow 0$, they are all proportional to the constant $\sigma_{\textsc{q}}$ associated with the translationally invariant QFT.    At stronger $u$,   $\sigma_{\textsc{q}1,2,3,4}$ may be complicated (spatially-averaged) correlations between $\mathcal{Q}$ and $\sigma_{\textsc{q}}$:  see (\ref{saa}), (\ref{sq2}) and (\ref{eq125}).   

We have written (\ref{eq2}) in the manner we did to emphasize asymptotic behavior.   When $u\ll \mathcal{Q}_0$, $\sigma \sim u^{-2}$ if $\mathcal{Q}_0 \ne 0$,  and this is a direct consequence of the fact that currents can only decay due to the very slow relaxation of momentum, as in a translationally invariant (momentum conserving) system at finite $\mathcal{Q}_0$, Galilean invariance imposes $\sigma=\infty$.\footnote{We can maintain an electric current without adding any energy by simply shifting to a moving reference frame.}     In contrast, when $u\gg \mathcal{Q}_0$, $\sigma$ is sensitive to typical behavior of local $\sigma_{\textsc{q}}$ and is not parametrically larger.     Remarkably, $\sigma_{\textsc{q}1}>0$ in any system where the local quantum critical conductivity never vanishes, so any such system is provably a conductor.   The physical intuition behind this is that a current can always flow locally due to finite $\sigma_{\textsc{q}}$, and so if these local currents can always flow, a global current flow can necessarily be established.   The upper bound is simply a statement that (up to subtleties involving conservation laws) we can bound the power dissipated (and accordingly the conductance), with the average electric field fixed, by assuming that the electric field within the system is uniform.   

When $u\ll \mathcal{Q}_0$, we can make contact with the memory matrix formalism and exactly compute $\sigma$ perturbatively, as we discuss in Section \ref{sec3}, and so in this regime we can do better than the bounds in (\ref{eq2}).  Indeed, in this regime, we can identify $\tau/\mathcal{M} \approx \sigma_{\textsc{q}}/u^2$, and so our bounds justify (\ref{drude}) in our class of hydrodynamic models.  Analogous phenomena may be responsible for the finite conductivity of mean field holographic models at all ``disorder strengths".

\begin{figure}[t]
\centering
\includegraphics[width=3.5in]{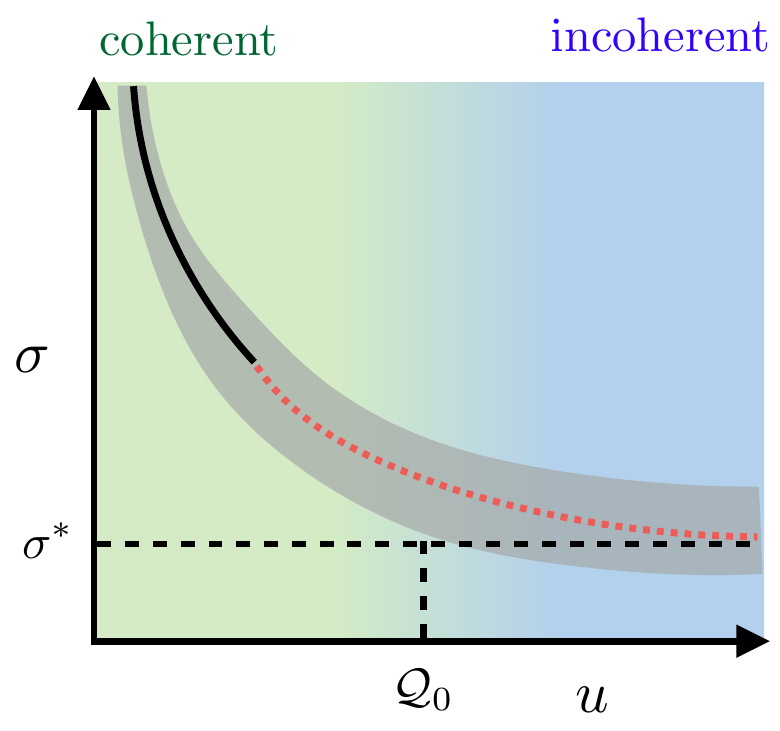}
\caption{A qualitative sketch of the coherent-incoherent transition realizable in our framework.  $\sigma$ denotes the value of a transport coefficient, such as electrical conductivity, and $u$ denotes the ``strength of randomness".   The solid black line shows our perturbative analytic computation of $\sigma \sim u^{-2}$ as $u\rightarrow 0$.   The dashed red line is the qualitative prediction of mean field models that $\sigma$ saturates at a finite value at strong disorder in a theory with quantum critical transport; in particular, $\sigma\ge \sigma^*$.   The gray shaded region corresponds to the region of $\sigma$ allowed by variational bounds on $\sigma$, in generic agreement with mean field models.  $u\sim \mathcal{Q}_0$ is the scale of the crossover between a coherent and incoherent metal.  }
\label{fig1}
\end{figure}

A pictorial summary of (\ref{eq2}) is shown in Figure \ref{fig1}, and the main quantitative result of this paper is the justification for Figure \ref{fig1} without any mean-field treatment of disorder, and the development of  new techniques to address the strongly disordered regime.  

After we had completed this work, \cite{donos1506, donos1507} appeared, and has  some overlap with ideas in Section \ref{sec4}.

\subsection{Motivation:  Beyond Resistor Lattices}
Though the main quantitative focus of this work is a set of computational tools to study hydrodynamic transport in relativistic fluids, such as in holography,  we also emphasize that the framework we are developing (with suitable generalizations) is sensible for a description of transport in strongly interacting condensed matter systems, without any reference to holography.     

A common approximation made in condensed matter is what we will refer to below as the ``resistor lattice" approximation, which in physical terms is the statement that the slow, hydrodynamic sector of the theory consists of only a conserved charge.   One may then model the emergent hydrodynamics -- a simple diffusion equation for charge -- as a local resistor network: see e.g. \cite{ruzin, halperin}.   As mentioned before, this is sensible if electrons scatter more frequently from impurities than they do from each other.

However, we will point out in Section \ref{sec32} that this approximation fails in a clean hydrodynamic system:  the necessary resistor lattice becomes nonlocal.   This is not a surprise.  What this paper clarifies is the technique that unifies the computation of transport in a weakly disordered (memory matrix) regime and a strongly disordered regime.  In doing so, we generalize  well-known variational techniques from resistor networks to account for convective transport.  In very special cases \cite{andreev} performed similar calculations, though did not elucidate the connections with the memory matrix formalism, or with resistor lattice technologies, which we generalize directly in the continuum in this paper.  Such resistor lattice methods -- commonly with an additional approximation called effective medium theory \cite{landauer, kirkpatrick}  -- have been used recently to study transport in a variety of experimentally realizable systems \cite{meir, sarma, demler}.    Our approach can generalize these computations to the regime when disorder is weak, and may result in interesting new experimental predictions.

We emphasize that the calculations in \cite{meir, sarma, demler} typically include non-relativistic effects such as Coulomb screening, or additionally approximate that electron-hole recombination is slow enough that both the electron and hole densities are hydrodynamic quantities.   We will not make either assumption in this paper, but the general framework and many computational methods we develop almost certainly extend quite naturally to account for these effects.

\section{Steady-State Hydrodynamics}\label{sec2}
We consider a strongly coupled QFT in $d$ spatial dimensions at finite temperature and density, on a flat spacetime.   It is necessary to generalize to curved spaces to connect with the results of Section \ref{sec4}, but  every result in this paper generalizes in the obvious way (replacing partial derivatives with covariant derivatives, $\int \mathrm{d}^d\mathbf{x} \rightarrow \int \mathrm{d}^d\mathbf{x} \sqrt{g}$, etc.), and so we will not do so explicitly for ease of presentation.   Without quasiparticles, the long time dynamics are that of charge, energy and momentum.   In this section, we will work with relativistic notation, though the techniques work for non-relativistic theories as well.   We focus on theories with a single conserved charge, but the techniques straightforwardly generalize to theories with multiple conserved charges.      Note that we will work in units with $\hbar=1$. 

We deform the microscopic Hamiltonian $H$  by an external chemical potential: \begin{equation}
H \rightarrow H - \int \mathrm{d}^{d}\mathbf{x} \; \bar A_\mu J^\mu .
\end{equation} with $J^\mu$ a conserved electrical current, $\bar F=\mathrm{d}\bar A$,  and\begin{equation}
\bar A = \bar \mu(\mathbf{x}) \mathrm{d}t,
\end{equation}
so if $\mathcal{Q}(\mathbf{x})$ is the local charge density operator, \begin{equation}
H \rightarrow H - \int \mathrm{d}^d\mathbf{x} \; \bar\mu(\mathbf{x}) \mathcal{Q}(\mathbf{x}).
\end{equation}The chemical potential in the fluid is thus $\bar \mu$.   We also assume  that the temperature is uniformly $T$, and that there is no fluid velocity,  in our background state.   This forms the basis of a consistent solution to  hydrodynamic equations, driven by the coupling $\bar\mu$ to an external bath, as we will derive below.   The steady-state hydrodynamic equations read (in relativistic notation) \cite{hkms} \begin{subequations}\label{hydroeq}\begin{align}
\partial_i T^{i\mu} &= \bar F^{\mu\nu}J_\nu, \\
\partial_i J^i &= 0, 
\end{align}\end{subequations}
where Greek indices denote spacetime indices and Latin indices denote spatial indices and $T^{\mu\nu}$ is the energy-momentum current.   We have implicitly taken expectation values over all operators in (\ref{hydroeq}) and will do so for the remainder of the paper.   Because we have sourced disorder in our fluid entirely through $\bar\mu(\mathbf{x})$, we do not need to couple any other dynamical sectors to the theory, though we will point out how this may be done perturbatively in Section \ref{sec:scalar}, when additional scalars contribute to disorder.     The coupling of the fluid to an external chemical potential means that both energy and momentum may be exchanged with the external bath.   

\begin{figure}
\centering
\includegraphics[width=5in]{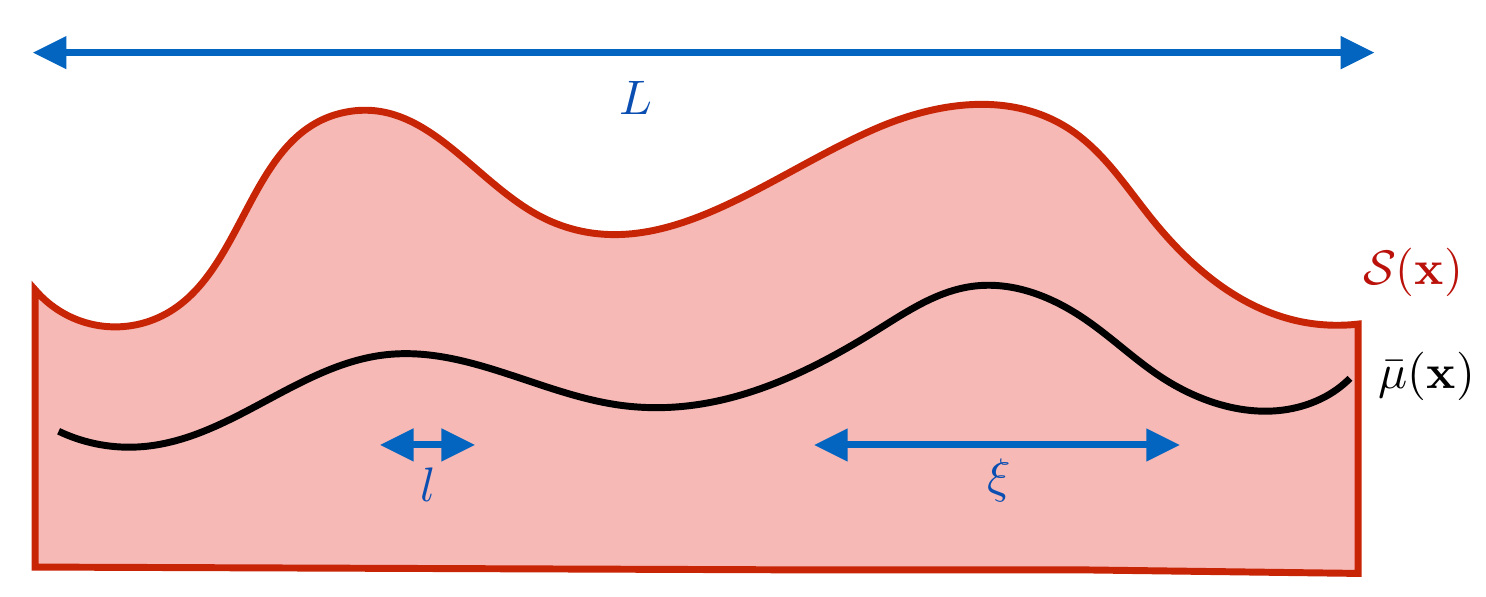}
\caption{We employ a separation of 3 length scales in this paper.   $\bar\mu$, and the local fluid properties such as entropy density $\mathcal{S}$, may vary substantially over the distance scale $\xi$.   We require $l \ll \xi$ for a hydrodynamic description to be sensible.   We will often put our fluids in a large but finite box of length $L\gg \xi$ as well.}
\label{figfluid}
\end{figure}

In order for hydrodynamics to be valid, it is necessary that $\bar\mu$ vary slowly in space, on a length scale $\xi$ which is large compared to the (possibly position-dependent) mean free path of the fluid $l$.       In our strongly interacting fluid, $l$ is the analogue of the electron-electron scattering length in traditional solid-state physics.   Without quasiparticles, it is best interpreted as the minimal length scale at which a hydrodynamic description is sensible.   The requirement that $\bar\mu$ vary slowly is often written as \begin{equation}
\left|\frac{\partial_x \bar\mu}{\bar\mu}\right| \ll \frac{1}{l},  \label{eq5}
\end{equation}though this should not be taken literally ($\bar \mu$ may vary slowly through $\bar \mu=0$).   The requirement we will assume henceforth in calculations is that, in Fourier space, $\bar\mu(\mathbf{k})$ is only non-negligible for $|\mathbf{k}|\xi \lesssim 1$.    It is not necessary that $\bar \mu$ be approximately the same at all points at space:\footnote{$|\bar\mu(\mathbf{x}_1)-\bar\mu(\mathbf{x}_2)|$ can be comparable to, or larger than,  $|\bar\mu(\mathbf{x}_1)|$, so long as $|\mathbf{x}_1-\mathbf{x}_2|\gg l$.}   disorder can be non-perturbative, with hydrodynamic coefficients such as viscosity and charge density, contained within $T^{\mu\nu}$ and $J^\mu$  in (\ref{hydroeq}), varying substantially over distances large compared to $l$; see Figure \ref{figfluid}.   This was noted in \cite{andreev} as well.   

In a quantum critical theory of dynamical exponent $z$,  one finds \begin{equation}
l \sim T^{-1/z}  \label{lT1z}
\end{equation} by dimensional analysis \cite{sachdev}.   Note that (\ref{eq5}) does not hold as $T\rightarrow 0$ -- it is thus important that we are considering the finite temperature response of the QFT.    (\ref{lT1z}) may be modified in models where the hydrodynamic limit can persist in regimes where $\bar\mu \gg T$, such as in holography \cite{davisoncold} (in this particular model one seems to find $l\sim \bar\mu^{-1}$), or when the expectation value of neutral scalar fields is large.   Explicit computations of $l$ are beyond the scope of this paper but are necessary to properly understand the regime of validity of hydrodynamics.    A conservative requirement is certainly to fix the background temperature $T$ to be uniform and large enough that $\xi \gg T^{-1/z}$ at all points in space, but this may be too strict, as we will see in holographic models.    

More liberally, one could only require that $\xi \gg l$ hold locally, with short wavelength disorder in ``hot" regions of space with small $l$,  and long wavelength disorder in ``cold" regions of space with large $l$.   So long as the solution to the hydrodynamic equations of motion itself varies slowly in the cold regions with large $l$,  then our hydrodynamic formalism should be an acceptable description of transport.

When (\ref{eq5}) holds, it is a sensible approximation (and standard in condensed matter physics) to assume that  thermodynamic and hydrodynamic coefficients, such as viscosity $\eta$ or charge density $\mathcal{Q}$, are \emph{local} and depend only on $\bar \mu(\mathbf{x})$.   We then -- very crudely speaking -- put together pieces of homogeneous fluid of width $\xi$, whose equations of state are translation invariant, and smooth over the fluctuations from piece to piece.    Our approach to transport is to focus on the response of the low energy, hydrodynamic degrees of freedom, exactly treating their evolution across the slowly varying background fluid, as we now detail.   

For holographic theories, we do \emph{not} need to make the assumption $\xi \gg l$, or the assumption that all transport coefficients are functions of $\bar\mu$ alone.   It is remarkable that the mathematical framework we develop in this paper is nonetheless applicable to so many holographic computations.

\subsection{Linear Response:  A Warm-Up}
Let us begin with some simple calculations to get an intuitive feeling for hydrodynamic transport.  We work with first order hydrodynamics, and will justify this later in the section.   We will also assume that our theory is isotropic, another assumption which we relax later.

A first simple case to consider is when the only slow dynamics in the system are of charge.  As we mentioned previously, in this case the dynamics reduce to the solution of a diffusion equation: \begin{equation}
\partial_i \mdelta J_i = \partial_i \left(\sigma^{\mathrm{loc}}(\mdelta E_i-\partial_i \mdelta \tilde\mu)\right) = 0,  \label{eqjsimple}
\end{equation}
where $\mdelta E_i$ is the infinitesimal, constant, externally applied electric field, $\mdelta J_i$ is the infinitesimal electric current, and $\mdelta \tilde\mu$ is the infinitesimal local chemical potential, excited in response to $\mdelta E_j$.    $\sigma^{\mathrm{loc}}(\mathbf{x})$ is a transport coefficient which is inhomogeneous in space, and can be interpreted as the local conductivity of the theory.   This approximation is well known in condensed matter physics, and we mentioned it in the introduction.   Note that chemical potential gradients are equivalent to electric fields in the hydrodynamic equations of motion.    The electrical conductivity of such a system is defined as \begin{equation}
\mathbb{E}[\mdelta J_i] =\sigma_{ij}\mdelta E_j, 
\end{equation}
where $\mdelta J_i$ is evaluated on the unique solution to (\ref{eqjsimple}) where $\mdelta \mu$ obeys sensible boundary conditions (e.g., periodicity when disorder is periodic with period $L$), and we have denoted with $\mathbb{E}[\cdots]$ a uniform spatial average.   If the disorder is isotropic, then $\sigma_{ij} = \sigma \mdelta_{ij}$.    Note that $\sigma$ is not equivalent to $\mathbb{E}[\sigma^{\mathrm{loc}}]$.   Henceforth, we will define \begin{equation}
\mdelta \mu \equiv \mdelta \tilde\mu - x_j \mdelta E_j,
\end{equation}so that (\ref{eqjsimple}) can be written compactly as \begin{equation}
\partial_i \mdelta J_i = -\partial_i \left(\sigma^{\mathrm{loc}}\partial_i \mdelta \mu\right).
\end{equation}

Let us now account for convective transport -- this means that momentum is a long lived quantity and must be included in hydrodynamics.   If we neglect thermal transport, then we must modify (\ref{eqjsimple}) to account for the convective contributions to charge: \begin{equation}
\partial_i \mdelta J_i = \partial_i \left(\mathcal{Q}\mdelta v_i-\sigma^{\textsc{q}}\partial_i \mdelta \mu\right) = 0,
\end{equation}
where $\mathcal{Q}$ is the charge density, and $\mdelta v_i$ is the velocity field in the fluid.    $\sigma^{\textsc{q}} \ne \sigma^{\mathrm{loc}}$ is a ``quantum critical" transport coefficient corresponding to the flow of a current in the absence of any velocity field \cite{hkms}.   The momentum conservation equation allows us to determine $\mdelta v_i$, and we will show more carefully below that this equation is the following analogue of the Navier-Stokes equation: \begin{equation}
\mathcal{Q} \partial_i \mdelta \mu  = \partial_j \left( \eta \partial_j \mdelta v_i + \eta \partial_i\mdelta v_j+ \mdelta_{ij}\left(\zeta - \frac{2\eta}{d}\right)\partial_k \mdelta v_k \right)
\end{equation}
with $\eta$ the shear viscosity and $\zeta$ the bulk viscosity.   As before, $\mathcal{Q}$, $\sigma^{\textsc{q}}$, $\zeta$ and $\eta$ can all depend on position $\mathbf{x}$,  though not in an arbitrary way.  As we discussed previously, the $\mathbf{x}$-dependence of all these coefficients is fixed by their dependence on $\bar\mu(\mathbf{x})$, as determined in a locally homogeneous fluid.

\subsection{Linear Response:  Complete Theory}
Let us now describe the complete linear response theory which includes the response of  temperature, chemical potential and velocity to external fields.

About our background fluid we perturb the system with an infinitesimal electric field \begin{equation}
\mdelta E_i  =  \mathbb{E}\left[- \partial_i \mdelta \mu\right],
\end{equation}and an infinitesimal temperature gradient \begin{equation}
\mdelta \zeta_i  =  \mathbb{E}\left[-\frac{1}{T} \partial_i \mdelta  T\right].
\end{equation} Defining the heat current\begin{equation}
 Q^i \equiv T^{it}- \bar\mu J^i,
\end{equation}
we find that the heat current is conserved (divergenceless): \begin{equation}
\partial_i Q^i = \partial_i T^{it} - \partial_i (\bar \mu J^i) = - \bar F_{ti}J_i -J_i \partial_i \bar \mu = 0.
\end{equation} 
The thermoelectric response of the theory is given by the matrices: \begin{equation}
\left(\begin{array}{c} \mathbb{E}[ \mdelta J_i] \\ \mathbb{E}[\mdelta Q_i] \end{array}\right) = \left(\begin{array}{cc} \sigma_{ij} &\ \alpha_{ij} T \\ \bar\alpha_{ij} T &\ \bar\kappa_{ij} T\end{array}\right) \left(\begin{array}{c} \mdelta E_j \\ \mdelta \zeta_j \end{array}\right).   \label{jq}
\end{equation} 

Let us define \begin{subequations}\begin{align}
\mdelta \Phi^\alpha &\equiv \left(\begin{array}{c} \mdelta \mu \\ T^{-1} \mdelta T\end{array}\right), \\
\mdelta F^\alpha_i &\equiv \left(\begin{array}{c} \mdelta E_i \\ \mdelta \zeta_i \end{array}\right), \\
\mdelta \mathcal{J}^\alpha_i &\equiv \left(\begin{array}{c} \mdelta J_i \\ \mdelta Q_i \end{array}\right), 
\end{align}\end{subequations}where the $\alpha$ vector index denotes charge (q) or heat (h).   Note that we may write bold-face vectors below, but this always refers to spatial indices only -- we will always write out the $\alpha$ index explicitly in equations.  We then may write (\ref{jq}) as \begin{equation}
\mathbb{E}[\mdelta \mathcal{J}^\alpha_i] = \sigma^{\alpha\beta}_{ij} \mdelta F^\beta_j.
\end{equation}

We write down the gradient expansion of hydrodynamics to first order in derivatives acting on $\mdelta T$, $\mdelta \mu$ and $\mdelta v_i$, by expanding stress tensor and charge current in terms of the linear response $\mdelta T$, $\mdelta \mu$  and $\mdelta v_i$ of the fluid.     The charge and heat conservation equations of the fluid may be written as \begin{subequations}\begin{align}
0 &= \partial_i \left(\mathcal{Q}\mdelta v_i - \sigma^{\textsc{q}}_{ij}\partial_j \mdelta\mu - \alpha^{\textsc{q}}_{ij}\partial_j \mdelta T\right), \\
0 &= \partial_i \left(T\mathcal{S}\mdelta v_i - T \bar\alpha^{\textsc{q}}_{ij}\partial_j \mdelta\mu - \bar\kappa^{\textsc{q}}_{ij}\partial_j \mdelta T\right),
\end{align}\end{subequations} which we henceforth package into the more compact form \begin{equation}
0 = \partial_i  \mdelta \mathcal{J}^\alpha_i = \partial_i \left[\rho^\alpha \mdelta v_i - \Sigma^{\alpha\beta}_{ij} \partial_j \mdelta\Phi^\beta\right],  \label{maineq1}
\end{equation}
where \begin{equation}
 \rho^\alpha \equiv \left(\begin{array}{c} \mathcal{Q} \\  T\mathcal{S} \end{array}\right),
\end{equation}
where $\mathcal{Q}$ is the electric charge density and $\mathcal{S}$ is the entropy density.   $\Sigma^{\alpha\beta}_{ij}$ correspond to diffusive transport coefficients that couple charge and heat flows to gradients in $\mdelta \mu$ and $\mdelta T$, even in the absence of any convective (non-vanishing $v_i$) fluid motion.   In particular, $\Sigma^{\text{qq}}_{ij} = \sigma^{\textsc{q}}_{ij}$ corresponds to ``quantum critical" conductivity and is typically assumed to vanish in a non-relativistic theory without particle-hole symmetry, as in \cite{andreev}.    $\Sigma^{\mathrm{qh}}_{ij} = \alpha^{\textsc{q}}_{ij}$ corresponds to an intrinsic diffusive conductivity that couples charge and heat flows.   In standard non-relativistic theories, only $\Sigma^{\mathrm{hh}}_{ij} = \bar\kappa^{\textsc{q}}_{ij}$ is nonvanishing, as in \cite{andreev}.  All three are non-zero in relativistic systems \cite{hkms}.  We assume that \begin{equation}
\Sigma^{\alpha\beta}_{ij} = \Sigma^{\beta\alpha}_{ji},  \label{diffsym1}
\end{equation}and that locally $\Sigma$ be a positive definite matrix (note $\alpha i$ and $\beta j$ group together for purposes of matrix inversion).   This is sensible from the point of view of the second law of thermodynamics.  Indeed, in isotropic theories, the second law provides more constraints on these transport coefficients than (\ref{diffsym1}) alone \cite{hkms}, but we can and will relax these constraints in the technical formalism we develop without substantially altering any physical content.   This loose treatment of entropic constraints proves useful in Section \ref{sec4}.

The momentum conservation equation becomes\footnote{Note that $\mdelta\mathcal{Q} = (\partial \mathcal{Q}/\partial \mu) \mdelta \mu + (\partial \mathcal{Q}/\partial T)\mdelta T$.}  \begin{align}
 \partial_i \mdelta P +  \partial_j \mdelta \mathcal{T}_{ij} &=\mdelta \left[\mathcal{S}\partial_i T + \mathcal{Q}\partial_i \mu \right]+  \partial_j \mdelta \mathcal{T}_{ij}=  \partial_i \mdelta  \left( \Phi^\alpha \rho^\alpha \right)  - \partial_j \left[ \eta_{ijkl}  \partial_l \mdelta v_k\right] = \mdelta \mathcal{Q} \partial_i \bar\mu,
\end{align}
 where $\mathcal{T}_{ij}$ is the viscous stress tensor, $P$ is the pressure, $\eta_{ijkl}$ is the viscosity tensor with symmetries \begin{equation}
 \eta_{ijkl}= \eta_{jikl} = \eta_{ijlk} = \eta_{klij},  \label{diffsym2}
 \end{equation} and we have used the fact that thermodynamic relations imply \begin{equation}
 \partial_i P = \mathcal{S} \partial_i T + \mathcal{Q} \partial_i \mu,  \label{dip}
 \end{equation} Now since $T$ is constant on the background, and as the background $\mu$ is simply given by $\bar \mu$,  we cancel the two $\mdelta \mathcal{Q}$ terms,and we are left with \begin{equation}
 0 = \rho^\alpha\partial_i \mdelta\Phi^\alpha  - \partial_j \left[ \eta_{ijkl}  \partial_k \mdelta v_l\right].  \label{maineq2}
 \end{equation}
 In the above equations, $\rho$, $\Sigma$ and $\eta$ are all smooth functions of $\bar\mu(\mathbf{x})$, varying on large length scales compared to $l$.

 (\ref{dip}), along with (\ref{hydroeq}) and the fact that $\mathcal{J}^\alpha_i = \mathcal{T}_{ij}=0$ on the background solution, demonstrates that the background solution to the hydrodynamic equations indeed exists.    $\mathcal{J}^\alpha_i=0$ on the background because the hydrodynamic equations only depend on $\bar\mu - \mu$, which identically vanishes \cite{hkms}.    Though we expect that disorder implies that $\rho^\alpha$, $\Sigma^{\alpha\beta}_{ij}$ and $\eta_{ijkl}$ are all functions of $\bar\mu$ alone, we will not comment further the precise nature of this dependence, and a microscopic computation is necessary in general.
  
(\ref{maineq1}) and (\ref{maineq2}) are linear and have a unique solution when subject to appropriate boundary conditions.   These boundary conditions will be periodic boundary conditions in a large box of size $L$ in all directions, up to non-trivial gradients $\mdelta F^\alpha_i = \mathbb{E}[-\partial_i \mdelta \Phi^\alpha]$.    We also stress that $\mdelta\Phi^\alpha$ only enters the equations of motion through derivatives -- this is crucial in order for the linear response problem to be well posed on spaces that are periodic or compact.   
Henceforth we will drop the $\mdelta$ so as to avoid clutter, with a few exceptions.

Having imposed these boundary conditions, we prove in Appendix \ref{apponsager} that, for any hydrodynamic transport computation,\begin{equation}
\sigma^{\alpha\beta}_{ij} = \sigma^{\beta\alpha}_{ji}.
\end{equation}
This is referred to as Onsager reciprocity, and is a non-trivial consistency check on this framework.    Note that this condition is violated when time-reversal symmetry (in the microscopic Hamiltonian $H$) is broken, e.g. by a background magnetic field \cite{hkms}.   We do not consider this possibility in this paper.

 As mentioned previously, we have truncated the hydrodynamic gradient expansion at first order.   Let us give some sensible, though non-rigorous, justifications for this.  The hydrodynamic gradient expansion can be organized as follows: \begin{subequations}\begin{align}
 \mathcal{T}_{ij} &\sim l \mathcal{T}^{(1)}_{ij} + l^2 \mathcal{T}^{(2)}_{ij} + \cdots, \\
 \mathcal{J}^\alpha_i - \rho^\alpha v_i &\sim    l \mathcal{J}^{(1)\alpha}_i+  l^2 \mathcal{J}^{(2)\alpha}_i + \cdots
 \end{align}\end{subequations}    
 $\mathcal{T}^{(n)}_{ij}$ corresponds to the coefficient of the stress tensor carrying $n$ spatial derivatives; similarly for $\mathcal{J}^{(n)\alpha}_i$.     This is a qualitative statement -- the basic idea is that the coefficients of $\mathcal{T}_{ij} \sim  l^n \epsilon/v$ at $n^{\mathrm{th}}$ order in derivatives, e.g., with $\epsilon$ the energy density and $v$ a velocity scale such as the speed of sound, and so we have extracted out the overall scaling in $l$ above.   Assuming that the solution $\Phi$ and $v_i$  varies over the length scale $\xi \gg l$,    we see that higher derivative corrections to the charge, heat and momentum currents are suppressed by powers of $l/\xi$, and thus can be neglected.    In the special case where diffusive charge and heat transport dominates, this argument can be made rigorous.    When the convective contributions cannot be ignored, this argument is not rigorous -- not all coefficients $\rho$, $\Sigma$ and $\eta$ scale as the same power of $l$, in general, and so it is not obvious to prove that $\Phi$ and $v_i$ must vary on the length scale $\xi$.   However, this is still a plausible assumption -- any rapidly oscillatory $\Phi$ and $v_i$ on short length scales compared to $\xi$ seems non-sensible as a static solution, since static solutions to dissipative hydrodynamics tend to be ``as close as possible" to equilibrium, given the boundary conditions;  the variational methods we will develop in this paper also suggest that it is unlikely to have fast variations of $\Phi$ and $v_i$.    This general framework readily generalizes to account for higher derivative corrections to hydrodynamics, if one wishes to directly include them, but we will not include them in this paper.    (\ref{maineq1}) and (\ref{maineq2}) are not well-posed until we include first order corrections to hydrodynamics,  so it is necessary to work at least to this order in the gradient expansion.
 
In the absence of other dynamical sectors of the theory, it is necessary that either $\mathcal{S}$ or $\mathcal{Q}$ be position dependent in order to obtain finite thermoelectric conductivities.  Indeed, if both $\mathcal{S}$ and $\mathcal{Q}$ are constants, there is a zero mode in (\ref{maineq1}) and (\ref{maineq2}) corresponding to uniform shifts in $v_i$ and $\mathcal{J}^\alpha_i$.   This zero mode is  responsible for infinite dc transport coefficients in a fully translation invariant theory.   Mathematically, we could break translation invariance only in $\Sigma$ or $\eta$ and still have this zero mode.  However, in a microscopic theory $\Sigma$, $\eta$, $\mathcal{S}$ and $\mathcal{Q}$ are not arbitrary but are fixed by equations of state that relate these parameters to $\bar\mu$, so in general both will be inhomogeneous.     Alternatively, as we will discuss in Section \ref{sec:scalar}, it is possible to add other dynamical disordered sectors of the theory which lead to finite conductivities even when $\mathcal{S}$ and $\mathcal{Q}$ are constants.

Let us also briefly mention the issue of momentum relaxation times.  In many holographic mean field models of disorder, the momentum relaxation time can be parametrically fast \cite{gouteraux2}.   In these hydrodynamic models, the ``momentum relaxation time" is parametrically slower than the mean free time ($1/T$ in most quantum critical models).   We do not explicitly compute this momentum relaxation time, and a single momentum relaxation time will not be easily definable when disorder is non-perturbative.   We will see that it is possible to spoil coherent transport despite parametrically long lived local momentum currents.

It is also worth stressing that henceforth, when we refer to ``hydrodynamic" transport we refer to the transport equations being written in terms of (\ref{maineq1}) and (\ref{maineq2}).   Remarkably, essentially all of our results rely only on the structure of these equations being obeyed, and not on $\mathcal{S}$ and $\mathcal{Q}$ obeying thermodynamic Maxwell relations --  holographic horizon fluids do not obey any obvious Maxwell relations.    If one has a microscopic system of interest, with known equations of state, they may simply take the above equations and numerically solve them.   So the point of this paper is less to describe complicated (numerical) solutions to these equations of motion, but rather to elucidate simple and universal physical consequences of hydrodynamic transport:   first through exact results for weakly disordered theories, and then through a combination of rigorous bounds and heuristic arguments for strongly disordered theories.  Numerical solutions to these equations, and a discussion of their relevance to realistic quantum critical systems, will be presented elsewhere.

\section{Weak Disorder Limit}\label{sec3}
In this section, we specialize to the weak disorder limit in which slow momentum relaxation dominates the conductivities.  In this limit we can make direct contact with the memory matrix formalism \cite{zwanzig, mori, forster}, and provide physically transparent derivations of many previous results derived within this formalism, in the overlapping regime of validity of hydrodynamics and the memory function formalism.    We simply quote the results of this approach (see e.g. \cite{lucasMM}):  if the Hamiltonian of our weakly disordered system is \begin{equation}
H = H_0 - \int \mathrm{d}^d\mathbf{x}\; h(\mathbf{x})\mathcal{O}(\mathbf{x}).
\end{equation} with $H_0$ translation invariant, and $\mathcal{O}$ an operator in the theory coupled to the field $h$, then the memory matrix formalism predicts, at leading order in perturbation theory: \begin{equation}
\sigma^{\alpha\beta}_{ij} \approx \mathbb{E}\left[ \rho^\alpha 
\right]\mathbb{E}\left[\rho^\beta\right] \left[\sum_{\mathbf{k}} \; k_i k_j |h(\mathbf{k})|^2\left[ \lim_{\omega\rightarrow 0} \frac{\mathrm{Im}\left(G_{\mathcal{OO}}(\mathbf{k},\omega)\right)}{\omega}\right] \right]^{-1} \equiv \mathbb{E}\left[ \rho^\alpha 
\right]\mathbb{E}\left[\rho^\beta\right] \Gamma^{-1}_{ij} \label{eqmm}
\end{equation}
In some models of strange metals appropriate for real world modeling, some care is required in defining $\mathcal{Q}$ \cite{patel}.    Formally, the memory matrix formalism is exact, but the formalism does not appear to be tractable in practice beyond leading order, in higher dimensional models.

Let us briefly note our conventions in Fourier space.   Fourier transforms are defined as \begin{equation}
\mathcal{O}(\mathbf{k}) = \frac{1}{L^d} \int \mathrm{d}^d\mathbf{x} \; \mathrm{e}^{-\mathrm{i}\mathbf{k}\cdot\mathbf{x}}\mathcal{O}(\mathbf{x}).
\end{equation}
We will often assume that disordered sectors of the fluid have (zero-mean) Gaussian fluctuations: e.g., quenched disorder on $\mathcal{O}$ would scale as \begin{subequations}\begin{align}
\mathbb{E}_{\mathrm{d}}[\mathcal{O}(\mathbf{k})] &= 0, \\
\mathbb{E}_{\mathrm{d}}[\mathcal{O}(\mathbf{k})\mathcal{O}(\mathbf{q})] &=  \frac{V^2_{\mathcal{O}}}{N}\mdelta_{\mathbf{k},-\mathbf{q}}, \;\;\;\; (|\mathbf{k}|\xi \lesssim 1)
\end{align}\end{subequations}where $N\gg 1$ represents the number of Fourier modes which ``meaningfully contribute" to $\mathcal{O}(\mathbf{x})$, and $\mathbb{E}_{\mathrm{d}}$ denotes an average over quenched disorder: \begin{equation}
 N \sim \left(\frac{L}{\xi}\right)^d.
 \end{equation}These definitions are chosen so that $\mathbb{E}[\mathcal{O}(\mathbf{x})^2] = V_{\mathcal{O}}^2$.    We will use these conventions throughout the paper.   Such disorder is consistent with  (\ref{eq5}).  At a typical point in the fluid, \begin{equation}
 \left|\frac{\partial_x \mathcal{O}}{\mathcal{O}}\right|^2 \sim \frac{\mathbb{E}[(\partial_x \mathcal{O})^2]}{\mathbb{E}[\mathcal{O}^2]} \sim \frac{\xi^{-2} V_{\mathcal{O}}^2}{V_{\mathcal{O}}^2} \sim \frac{1}{\xi^2}.
 \end{equation}
 To obtain the numerator in the third step above, it is helpful to go to Fourier space, and note that $|\mathbf{k}| \lesssim \xi^{-1}$ for all non-negligible modes.

 In \cite{hkms} and many subsequent works, within the hydrodynamic approach to transport, the momentum transport equation is modified to \begin{equation}
\partial_j T^{ji} = -\frac{T^{ti}}{\tau} + \cdots,
\end{equation}where $T^{ti}$ is the momentum density, and $\tau$ is a phenomenological relaxation time that is subsequently computed using memory functions.\footnote{In contrast, \cite{lucasMM} has recently used the memory matrix formalism to recover hydrodynamic transport, with expressions for the phenomenological $\tau$ and other thermodynamic coefficients expressed in terms of microscopic Green's functions.   This is a more complete approach to the problem, but does not generalize easily to higher orders in perturbation theory.}   However, as we will see in this section, at least for dc transport, it is actually not necessary to add in $\tau$ by hand.   With weak disorder, the dc transport can be accounted for exactly from hydrodynamics, and (\ref{eqmm}) recovered provided that the equations of motion properly account for disorder.     

Interestingly, our hydrodynamic approach requires the disorder is always long wavelength compared to $l$, and  the memory matrix formalism does not require this restriction.  Nonetheless, at leading order in perturbation theory, we will recover the exact memory matrix formula for transport coefficients from hydrodynamic considerations.  It is also worth noting that the memory function approach is equivalent to holographic computations of transport in their overlapping regime of validity \cite{lucas}.   Thus, all three approaches give the same picture of transport, which is best physically understood in terms of this simple hydrodynamic framework (notwithstanding the regime of validity).

\subsection{Disorder Sourced by Scalar Operators}\label{sec:scalar}
Let us begin with the case where the operator $\mathcal{O}$ is a scalar field. We assume that all hydrodynamic coefficients are $\mathbf{x}$-independent -- $h$ is the only disordered parameter.  In this case, (\ref{maineq2}) must be modified:\begin{equation}
\rho^\alpha \partial_j \Phi^\alpha + \partial_i \mathcal{T}_{ij} =  \mdelta \mathcal{O} \partial_j h .  \label{maineq2scalar}
\end{equation}We place a $\mdelta$ on $\mathcal{O}$ to distinguish the response due to the electric field from the background.  The scalar's static equation of motion  is \cite{kadanoff} \begin{equation}
\int \mathrm{d}^d\mathbf{y}\; G_{\mathcal{OO}}^{-1}(\mathbf{x}-\mathbf{y},\omega=0)  \mathcal{O}(\mathbf{y}) = h(\mathbf{x}).
\end{equation}The Green's function $G_{\mathcal{OO}}$ is the retarded Green's function of the translationally invariant Hamiltonian $H_0$:  in position space,\begin{equation}
G_{\mathcal{OO}}(\mathbf{x},t) \equiv \mathrm{i}\mathrm{\Theta}(t) \langle [\mathcal{O}(\mathbf{x},t),\mathcal{O}(\mathbf{0},0)]\rangle
\end{equation}with the average $\langle \cdots \rangle$ taken over quantum and thermal fluctuations, and $\mathrm{\Theta}$ the Heaviside step function.  We emphasize that while $G_{\mathcal{OO}}$ is the true quantum Green's function of $\mathcal{O}$, and an intricate quantum mechanical computation may be necessary to compute it, $G_{\mathcal{OO}}$ does play the role of the coefficient of proportionality in the linear response of macroscopic, thermal expectation values.   

Let us make an ansatz for the solution to the hydrodynamic equations, and show that it is consistent with all conservation laws.   Our ansatz is that the only divergent terms in linear response, as $h\rightarrow 0$, are $ \mathbf{v} \sim h^{-2}$ and $\mdelta \mathcal{O} \sim h^{-1}$.   Furthermore, at leading order $ \mathbf{v}=\mathbf{v}_0$ is a constant.   All other spatially dependent response is $\mathrm{O}(h^0)$ and we will see that it can be neglected in the computation of $\sigma_{ij}$ at leading order.

The leading order response of $\mathcal{O}$ in the $h\rightarrow 0$ limit is best computed in the rest frame of the fluid, which has shifted.   It is simplest to Fourier transform to momentum space as well: \begin{equation}
\mathcal{O}(\mathbf{k},-\mathbf{k}\cdot\mathbf{v}_0)_{\mathrm{co-moving}} = G_{\mathcal{OO}}(\mathbf{k},  -\mathbf{k}\cdot\mathbf{v}_0) h(\mathbf{k},-\mathbf{k}\cdot\mathbf{v}_0)_{\mathrm{co-moving}}.
\end{equation}Here, everything is measured in the co-moving frame of the fluid, and so the only non-vanishing $h$ (and therefore $\mathcal{O}$) will have this special relation between $\mathbf{k}$ and $\omega$.  Note, of course, that $\mathcal{O}(\mathbf{k},-\mathbf{k}\cdot\mathbf{v}_0)_{\mathrm{co-moving}} = \mathcal{O}(\mathbf{k})$, as measured in the original rest frame, and similarly for $h(\mathbf{k})$.  We wish to keep only the linear response coefficient, proportional to $\mathbf{v}_0$:\footnote{Note that this linear term in perturbations is parametrically large in $h$ -- but it is linear in $F^\alpha_i$.   Our perturbative parameter is first and foremost $F^\alpha_i$, since we are computing a linear response transport coefficient.  And while the background may also be treated perturbatively in disorder strength, we must take $F^\alpha_i\rightarrow 0$ \emph{before} $h\rightarrow 0$.} \begin{equation}
\mdelta \mathcal{O}(\mathbf{k}) = -h(\mathbf{k})\frac{\partial G_{\mathcal{OO}}(\mathbf{k},0)}{\partial \omega} \mathbf{k}\cdot\mathbf{v}_0  =  -\mathrm{i}h(\mathbf{k}) \left[ \lim_{\omega\rightarrow 0} \frac{\mathrm{Im}\left(G_{\mathcal{OO}}(\mathbf{k},\omega)\right)}{\omega}\right] \mathbf{k}\cdot \mathbf{v}_0.
\end{equation}In the latter equality we have used reality propeties of Green's functions, and assumed analyticity near the real axis for $\mathbf{k}\ne \mathbf{0}$.    

Now, let us study the momentum conservation equation, averaged over space, so the derivatives of the stress tensor do not contribute: \begin{align}
0 &= \sum_{\mathbf{k}} \mdelta \mathcal{O}(\mathbf{k}) (-\mathrm{i}k_j h(-\mathbf{k}))+ \rho^\alpha F^\alpha_j \notag \\
&= -\sum_{\mathbf{k}} \; k_i k_j |h(\mathbf{k})|^2\left[ \lim_{\omega\rightarrow 0} \frac{\mathrm{Im}\left(G_{\mathcal{OO}}(\mathbf{k},\omega)\right)}{\omega}\right]   v_{0i} + \rho^\alpha F^\alpha_j = -\Gamma_{ij}v_{0j} + \rho^\alpha F^\alpha_i.
\end{align}
At leading order, the electric current is uniform:\begin{equation}
\mathcal{J}^\alpha_i \approx \rho^\alpha v_{0i} = \rho^\alpha \rho^\beta \Gamma^{-1}_{ij}F^\beta_j,  \label{jpv}
\end{equation} which gives us (\ref{eqmm}).    It is straightforward to generalize to the case where there are multiple types of scalar fields.

Let us now argue that the ansatz (and thus results) we have found are self-consistent.   If we do not average the momentum conservation equation over space, then the $\mdelta\mathcal{O}\partial_j h$ term is not translationally invariant, and this will induce corrections to $ T$, $ \mu$ and $ \mathbf{v}$ which are spatially varying.   However, $\mdelta\mathcal{O} \sim h^{-1}$ and so these spatially varying corrections will be $\sim h^0$.   Indeed, it is easy to see that (\ref{maineq1}) and (\ref{maineq2scalar}) are consistent with the leading order inhomogeneous response (except in $\mdelta \mathcal{O}$) arising at this subleading order.   Thus, our ansatz is indeed correct in the asymptotic limit $h\rightarrow 0$, and we have derived from hydrodynamic principles the momentum relaxation times derived via the memory function formalism.   The computation above  is completely analogous to the holographic computation of \cite{lucas}.

\subsection{Disorder Sourced by Chemical Potential}\label{sec32}
In this section, we consider the case where \begin{equation}
\bar\mu = \mu_0 + \epsilon \hat\mu,
\end{equation}with $\mathbb{E}[\hat\mu]=0$, and $\epsilon \ll 1$ a small perturbative parameter.   Alternatively, we can write \begin{equation}
\bar\Phi^\alpha = \Phi_0^\alpha + \epsilon \hat\Phi^\alpha
\end{equation}with $\bar\Phi^{\mathrm{h}}=T\mathcal{S}$ and $\hat\Phi^{\mathrm{h}}=0$ -- this will be more compact notation for subsequent manipulations.    We will denote $\hat\rho^\alpha$ with the fluctuations in the charge and entropy densities associated with $\hat\mu$ -- in general both will be non-zero.   For simplicity, we assume that the background fluid is isotropic, though the technique certainly generalizes (but with more cumbersome calculations).  \cite{dsz} studied similar problems employing hydrodynamic Green's functions in the memory matrix formalism.

Again, we make the ansatz that the only response at $\mathrm{O}(\epsilon^{-2})$ is a constant velocity field $\mathbf{v}_0$, so that $\mathcal{J}^\alpha_i$ is again approximated by (\ref{jpv}).    At $\mathrm{O}(\epsilon^{-1})$, there are $\mathbf{x}$-dependent corrections to $T$, $\mu$, and $\mathbf{v}$.   A similar calculation to before gives \begin{equation}
\Gamma_{ij} = \sum_{\mathbf{k}} \; k_ik_j \hat\rho^{\alpha}(-\mathbf{k}) \left[\frac{1}{\eta^\prime}\rho^\alpha_0 \rho^\beta_0 + k^2 \Sigma^{\alpha\beta}\right]^{-1}   \hat \rho^\beta(\mathbf{k}) \equiv\sum_{\mathbf{k}} \; k_ik_j \hat\rho^{\alpha}(-\mathbf{k}) (\mathfrak{m}^{-1})^{\alpha\beta}   \hat \rho^\beta(\mathbf{k}),   \label{gammaij32}
\end{equation}with \begin{equation}
\eta^\prime \equiv \eta \left(2-\frac{2}{d}\right) +\zeta,
\end{equation}
with $\eta$ the shear viscosity and $\zeta$ the bulk viscosity.   We provide more details in Appendix \ref{apppert}.

Let us briefly discuss some simplisitic limiting cases of (\ref{gammaij32}) and give some analytic insight into the solutions -- in general, the solutions will be more complicated than what we write here.   

First, let us begin with the case with $\eta^\prime \rightarrow \infty$ and $\hat\rho^{\mathrm{h}} \approx 0$.  Suppose further $\hat\rho^{\mathrm{q}}$ are Gaussian disordered random variables: \begin{subequations}\begin{align}
 \mathbb{E}_{\mathrm{d}}[\hat\rho^\alpha(\mathbf{k})] &= 0, \\
 \mathbb{E}_{\mathrm{d}}[\hat\rho^{\mathrm{q}}(\mathbf{k})\hat\rho^{\mathrm{q}}(\mathbf{q})] &= \frac{u^2}{N}\mdelta_{\mathbf{k},-\mathbf{q}}, 
 \end{align}\end{subequations}with $\mathbb{E}_{\mathrm{d}}[\cdots]$ denoting averages over the distribution of quenched disorder modes. Then we find \begin{equation}
 \mathbb{E}_{\mathrm{d}}[\Gamma_{ij}] \approx \frac{1}{Nd}\mdelta_{ij} \sum_{\mathbf{k}} \frac{u^2}{\Sigma^{\mathrm{qq}}} = \mdelta_{ij} \frac{u^2}{d\Sigma^{\mathrm{qq}}}  = \mdelta_{ij} \mathbb{E}\left[\frac{\hat{\mathcal{Q}}^2}{d\Sigma^{\mathrm{qq}}} \right].   \label{gamma40}
  \end{equation}
   Fluctuations of this quantity are suppressed in the limit $V_d\rightarrow\infty$ \cite{lucas1411}, so $\mathbb{E}_{\mathrm{d}}[\Gamma_{ij}] \approx \Gamma_{ij}$.  


An alternative simple case is thermal transport with $\mathcal{Q}=0$, and $\mathcal{S}\approx \mathcal{S}_0$ with small variations.  In this case, we may approximately neglect the $\Sigma$ contributions to $\mathfrak{m}$ as $\xi \rightarrow \infty$, and we find by a similar calculation to above: \begin{equation}
\mathbb{E}_{\mathrm{d}}[\Gamma_{ij}] \approx \mdelta_{ij} \mathbb{E}\left[\frac{\eta^\prime}{d} \frac{(\partial_i \mathcal{S})^2}{\mathcal{S}_0^2}\right],
\end{equation}
which is similar to results discussed in \cite{dsz}.    

To make contact between (\ref{gammaij32}) and the memory function framework, of course, we need to compute the retarded Green's functions for charge and heat.  Unfortunately, the Green's functions coupling charge and heat flow in a relativistic hydrodynamic system are quite messy \cite{kovtunlec}, and so we will prove the equivalence with (\ref{eqmm}) in a more abstract manner.    We proceed analogously to the previous case -- as above, we have showed that the leading order response of the fluid that contributes to $\mathcal{J}^\alpha_i$ at $\mathrm{O}(\epsilon^{-2})$ is from a constant shift to the velocity.   

The retarded Green's function is \emph{defined} as: \begin{equation}
\hat\rho^\alpha(\mathbf{k}) + \mdelta \rho^\alpha(\mathbf{k}) = G^{\alpha\beta}(\mathbf{k},-\mathbf{k}\cdot\mathbf{v}_0) \hat\Phi^\beta(\mathbf{k}) \approx \chi^{\alpha\beta}\hat\Phi^\beta(\mathbf{k}) - \mathbf{k}\cdot\mathbf{v}_0 \frac{\partial G^{\alpha\beta}(\mathbf{k},0)}{\partial \omega}  \hat\Phi^\beta(\mathbf{k}). 
\end{equation}
Of course, just as in Section \ref{sec:scalar}, we must use the boosted Green's function to obtain the linear response $\mdelta \Phi^\alpha$: the $\mathrm{O}(v^0)$ term is the response of the background fluid, and the $\mathrm{O}(v)$ term is the linear response contribution of interest for the computation of transport -- we focus on the latter ($\mdelta \rho^\alpha$) henceforth.     We can also relate $\mdelta \rho^\alpha$ to $\mdelta \Phi^\alpha$ by the static susceptibilities, since we are perturbing about a translationally invariant state: \begin{equation}
\mdelta \rho^\alpha = \chi^{\alpha\beta} \mdelta \Phi^\beta.
\end{equation}
Again, we can relate $\mathbf{v}_0$ to $\mathbf{F}^\alpha$ by spatially averaging (\ref{maineq2}): \begin{align}
\rho_0^\alpha F^\alpha_i &=  \sum_{\mathbf{k}} \; \hat \rho^\alpha(-\mathbf{k}) \mathrm{i}k_i \mdelta \Phi^\alpha(\mathbf{k}) =  \sum_{\mathbf{k}}  \; \hat \rho^\alpha(-\mathbf{k}) \mathrm{i}k_i \left(\chi^{-1}\right)^{\alpha\beta}\mdelta \rho^\beta \notag \\
&=  \sum_{\mathbf{k}} \; k_ik_j \hat\Phi^\alpha(\mathbf{-k}) \left[\lim_{\omega\rightarrow 0} \frac{\mathrm{Im}(G^{\alpha\beta}(\mathbf{k},\omega))}{\omega}\right]\hat\Phi^\beta(\mathbf{k}) v_{0j}.
\end{align}
It is straightforward to read off $\Gamma_{ij}$ from this equation, and we see that it agrees with the  generalization of (\ref{eqmm}) to multiple disordered quantities -- though of course at this point we use the fact that only $\hat\Phi^{\mathrm{q}} \ne 0$.    Since $\mathcal{J}^\alpha_i\approx \rho^\alpha v_{0i}$, we reproduce the results of the memory matrix formalism.   

We have worked through two specific examples of deriving (\ref{eqmm}) from hydrodynamics.   Of course one may need to generalize further, but it should be quite evident from the derivations above that the agreement between the memory matrix formalism and our hydrodynamic framework will persist.

It is possible to compute the transport coefficients at higher orders in perturbation theory, where the memory matrix formalism has become unwieldy enough that such a calculation has not yet been attempted.    Even at next order in perturbation theory, the corrections to the conductivity become quite messy.   We discuss the general structure of higher order computations in Appendix \ref{apppert}.   The key point is that  organizing the perturbative expansion in a hydrodynamic framework is straightforward in principle, albeit messy to carry out in practice.   This framework provides quantitative predictions for memory matrix calculations at higher orders in $\epsilon$, for classes of models perturbed by $\bar\mu$.

\subsubsection{Breakdown of the Resistor Lattice Approximation}
Finally, let us compare the results of this subsection with the ``resistor lattice" approximation:\begin{equation}
J_i \approx - \sigma_{ij}(\mathbf{x}) \partial_j \mu(\mathbf{x}),  \label{emteq}
\end{equation} with $\sigma_{ij} \ne \Sigma^{\mathrm{qq}}_{ij}$ taken to be a local function, determined in terms of local properties of the fluid, and and $-\partial_j \mu$ the local electric field in the sample.     Of course such a function $\sigma$ may be found by solving a linear algebra problem at each point in space, and so the question is whether this is a useful statement -- namely, whether $\sigma_{ij}(\mathbf{x})$ can be computed by appealing to local properties of the disordered QFT (on length scales large compared to $l$).   Essentially, can we integrate out $v_i$ and $T$, and be left with a local, dissipative description of electrical transport in terms of $\mu$ alone?   

  This is impossible in the weak disorder limit, though our comments appear to have broader validity whenever viscosity cannot be neglected.   $\partial_j \mu$ is actually inhomogeneous at leading order $\epsilon^{-1}$, and must be expressed in terms of a non-local integral over $\mathcal{Q}(\mathbf{x}) = \rho^{\mathrm{q}}(\mathbf{x}) \approx \chi^{\mathrm{qq}}\epsilon \hat\mu(\mathbf{x})$, with $\chi^{\mathrm{qq}}$ the charge-charge susceptibility, assumed to be constant at this order in perturbation theory.   This is derived in (\ref{eqphim1}), with the $X$ and $Y$ corrections in (\ref{eqphim1}) vanishing at leading order in $\epsilon$;  in this equation, the leading order behavior of $\mu$ is ``local in Fourier space", and becomes non-local in position space, in terms of the original disorder $\hat\mu$.    It is therefore generally impossible to find a function $\sigma_{ij} \sim \epsilon^{-1}$ expressable in terms of $\hat\mu$ or its (low order) derivatives, such that $J_i = \sigma_{ij}(-\partial_j \mu) = \text{constant} \sim \epsilon^{-2}$ (at leading order).

\section{Comparing to Holography}\label{sec4}\label{sec:holostripe}
Let us now compare with holographic results.   Many holographic results, valid in the weak disorder limit, are equivalent to the memory matrix results \cite{lucas} -- and therefore our hydrodynamic framework.   So our focus here will be on non-perturbative holographic results.

\begin{figure}[t]
\centering
\includegraphics[width=2.75in]{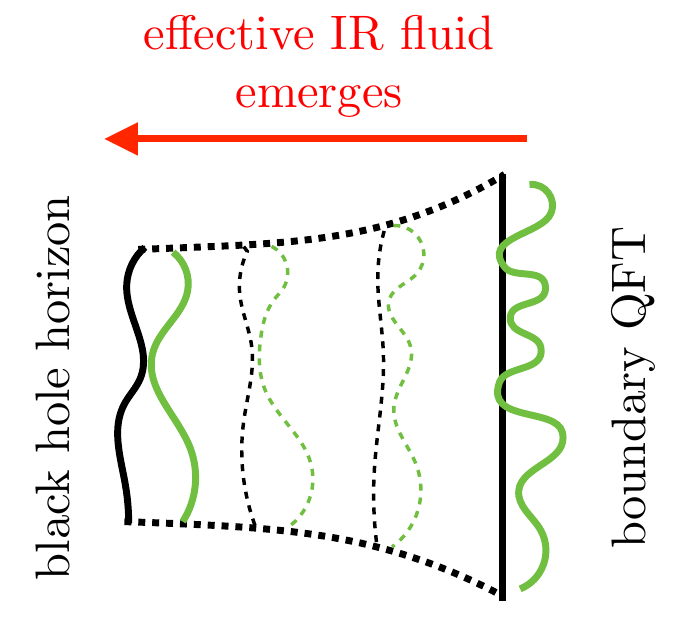}
\caption{A qualitative sketch of holography.  A finite temperature $T$ and density boundary theory is dual to an emergent gravitational theory in one extra spatial dimension (depicted above).    Strong disorder in the boundary theory (depicted in green) backreacts and leads to the formation of a lumpy charged black hole of Hawking temperature $T$.  The emergent black hole horizon is curved and is denoted in black.   The membrane paradigm suggests that dc transport can be computed in an emergent fluid living on the horizon, which can undergo renormalization relative to the ``bare fluid" in the boundary theory.}
\label{fig2}
\end{figure}

Our discussion of holography is brief -- for further details, consult the excellent reviews \cite{review1, review2, review3}.    Holography refers to a conjectured duality between a classical gravity theory in $d+2$ spacetime dimensions, in an emergent anti-de Sitter (AdS) space, and a strongly coupled QFT in $d$ spatial dimensions.   The strongly coupled QFTs (in every case where we know them explicitly) are large-$N$ matrix models, and can be thought to ``live" at the boundary of AdS.   Making the gravity theory classical is equivalent to sending the bulk Newton's gravitational constant to 0, and this makes bulk quantum gravity fluctuations negligible.   This corresponds to the limit $N\rightarrow \infty$ in the dual theory.   However, unlike vector models,   these matrix models do not behave at all like free theories, and encode rich quantum critical dynamics.  The  nonlinear dynamics of gravity is dual to the stress-energy sector of the boundary theory.   Furthermore, studying finite temperature dynamics becomes simply related to studying the dynamics in a black hole background.    These black holes will be assumed to have the same planar (or toroidal) topology of the boundary theory.    Adding a finite charge density in the boundary theory is dual to adding a bulk U(1) gauge field,  and charging the black hole under the associated U(1) charge.     Of interest for us in this paper is that holographic models can further be used to add strong disorder in addition to finite temperature and density.   We are interested in modeling disorder explicitly, and so the bulk geometry becomes inhomogeneous and rugged.   At small temperatures, the black hole gets pushed ``farther back" into the emergent bulk direction, and the bulk fields become significantly renormalized, with higher momentum modes usually decaying away, as depicted in Figure \ref{fig2}.

The precise duality allows us to compute correlation functions in our unknown QFT by solving gravitational equations instead.    The basic idea is as follows: correlation functions of the stress-energy tensor $T^{\mu\nu}$ and the U(1) current $J^\mu$ are respectively related to bulk Green's functions of the metric $g_{MN}$ and a gauge field $A_M$, all computed at (classical) tree level.   We are using $MN$ etc. to denote all coordinates, including the bulk radial coordinate.   For example, to compute the electrical conductivity, we add an explicit infinitesimal source for the bulk field $A_i$ at the AdS boundary, and then compute the expectation value of the current\footnote{This expectation value is also encoded near the boundary of AdS.} in the field theory in the background perturbed by $A_i$.

These computations are often intractable analytically.   However, in the simple case of dc transport, many analytic computations of dc transport in holography are performed using the membrane paradigm \cite{iqbal}.   In this case one can show that there is an analogous ``electric current" flowing on the black hole horizon, which is also conserved and whose average value equals that of $J^\mu$ in the boundary (an analogous more complicated story holds for the heat current).   The resulting computation of the conductivities then depend only on black hole horizon data.    It is natural to conjecture that we should solve the hydrodynamic transport problem using the equations of state of an emergent ``fluid" whose equation of state is related to local properties of the black hole horizon.   Indeed -- up to some subtleties we will see shortly -- this is the case.  

It is remarkable that  \emph{first order} hydrodynamics captures the transport problem on the emergent black hole horizons in holography, in all nonperturbative computations to date.    As we will see, however,  this emergent horizon fluid is not locally equivalent to a fluid in the boundary QFT -- instead it has undergone non-local renormalization, through the radial evolution of the bulk fields and geometry.       In addition, while we previously had to make assumptions that the disorder correlation length was large to justify our hydrodynamic formalism, no such justification is necessary (at least, a priori) for the holographic results to be valid.    When $\xi\rightarrow \infty$ in a holographic model, the differences between the horizon and boundary fluids become negligible, as was found in \cite{herzog}.

Let us begin with the striped models studied in \cite{donos1409},\footnote{Note that  their results simplify, in some special cases, to analytic results derived in \cite{chesler, peet}. The case where $\mathcal{Q}=0$ was also studied in \cite{ugajin}.} which consider gravitational solutions to the AdS-Einstein-Maxwell system: \begin{equation}
S = \int \mathrm{d}^4x\; \sqrt{-g} \left(R+6 - \frac{F^2}{4}\right).
\end{equation}
(we have set Newton's constant, the bulk charge, and the size of AdS equal to 1 to follow their notation).  Above $R$ is the Ricci scalar, and $F$ is the Maxwell tensor associated with the bulk gauge field.    In these models, translation invariance is only broken in the $x$ direction, but the boundary theory lives in $d=2$.  Let us summarize their results briefly.     They found that (after inverting their thermoelectric conductivity matrix):\begin{subequations}\label{donosdata1}\begin{align}
T\bar\kappa_{xx} = \sigma^{\mathrm{hh}}_{xx} &= \frac{16\mpi^2T^2}{X} \mathbb{E}\left[\mathrm{e}^{B}\right], \\
T\alpha_{xx} = \sigma^{\mathrm{qh}}_{xx} &=\frac{4\mpi T}{X} \mathbb{E}\left[ \mathrm{e}^{B}\frac{a_t}{H_{tt}}\right], \\
\sigma_{xx} = \sigma^{\mathrm{qq}}_{xx} &= \frac{1}{X} \mathbb{E}\left[ \mathrm{e}^{B}\left(\frac{a_t}{H_{tt}}\right)^2 + \frac{1}{S}\left(\partial_x B - \frac{\partial_xS}{S}\right)^2\right]
\end{align}\end{subequations}
where \begin{equation}
X = \mathbb{E}\left[ \mathrm{e}^{B}\left(\frac{a_t}{H_{tt}}\right)^2 + \frac{1}{S}\left(\partial_x B - \frac{\partial_xS}{S}\right)^2\right] \mathbb{E}\left[\mathrm{e}^{B}\right] - \mathbb{E}\left[ \mathrm{e}^{B}\frac{a_t}{H_{tt}}\right]^2
\end{equation} and $B$, $a_t$, $H_{tt}$ and $S$ are data associated with the solution of classical Einstein-Maxwell gravity, near the horizon of a black hole, as detailed below.    The near horizon geometry in their coordinate system was \begin{equation}\label{ds21}
\mathrm{d}s^2 \approx \frac{H_{tt}(x)}{4\mpi Tr} \mathrm{d}r^2 -4\mpi TrH_{tt}(x)\mathrm{d}t^2 +  S(x) \left[\mathrm{e}^{B(x)} \mathrm{d}x^2 + \mathrm{e}^{-B(x)}\mathrm{d}y^2\right], 
\end{equation}
with $r$ the radial coordinate, and $r=0$ denoting the location of the black hole.   The bulk gauge field only has a time-like component, whose value is $a_t$.   $T$  denotes the temperature of the boundary theory.

Let us postulate the following equations of state for the emergent  fluid on the horizon: \begin{subequations}\label{donosdata2}\begin{align}
\eta &= \frac{\mathcal{S}}{4\mpi} = S, \\
\mathcal{Q} &= \frac{Sa_t}{H_{tt}}, \\
\Sigma^{\mathrm{qq}} &= 1, \\
\Sigma^{\mathrm{qh}} &= \Sigma^{\mathrm{hh}} = 0.
\end{align}\end{subequations}
The first of these equations is the canonical result for $\eta/\mathcal{S}$ in a strongly coupled theory \cite{kss}, which also holds for the charged black holes here \cite{mas, kss2}, in the translationally invariant limit,  though this universal ratio can be different in mean-field disordered black holes \cite{gouteraux2}.   The last of these equations was argued to occur in holographic models in \cite{lucasMM}, by matching $\omega=0$ results of massive gravity.    More recently, it has been pointed out that this is not the correct interpretation of $\Sigma^{\alpha\beta}$ in the boundary theory, and this becomes discernable at finite $\omega$ \cite{davison15}.   However, we will see that this prescription can describe correctly an emergent fluid, associated with data on the horizon, whose hydrodynamic response is equivalent to (\ref{donosdata1}).  The inequivalence of the boundary fluid and this emergent ``horizon fluid" is an important subtlety, and one we will not resolve in this paper.

We also need to make two more assumptions.   The first is rather simple -- let us suppose (\ref{donosdata2}) is valid for the disordered model, with $x$-dependence trivially put in:  e.g., $\eta(x) = S(x)$.   This is in accordance with our logic in Section \ref{sec2}.  The second assumption is that the boundary fluid lives on a curved space with metric \begin{equation}
\mathrm{d}s^2 \equiv \gamma_{ij} \mathrm{d}x^i\mathrm{d}x^j = \mathrm{e}^{B(x)} \mathrm{d}x^2 + \mathrm{e}^{-B(x)}\mathrm{d}y^2,
\end{equation}which differs from (\ref{ds21}) by a conformal rescaling.   Intuitively this can be argued for on the grounds that $S$ determines $\mathcal{S}$ and $\eta$, and therefore should not determine the boundary metric $\gamma_{ij}$, since we expect that $\gamma_{ij}=\mdelta_{ij}$ in a translationally invariant isotropic model, even though $S\ne 1$ in general.    We compute the thermoelectric conductivties for such a fluid using special techniques for striped systems, discussed in Appendix \ref{appstripe}, and we find \begin{subequations}\label{donosresult}\begin{align}
\left(\sigma^{-1}\right)^{\mathrm{qq}}_{xx} &=  \mathbb{E}\left[\mathrm{e}^{B}\right], \\
\left(\sigma^{-1}\right)^{\mathrm{qh}}_{xx} &= -\frac{1}{4\mpi T}\mathbb{E}\left[ \mathrm{e}^{B}\frac{a_t}{H_{tt}}\right]  , \\
\left(\sigma^{-1}\right)^{\mathrm{hh}}_{xx} &= \frac{1}{(4\mpi T)^2}   \mathbb{E}\left[ \mathrm{e}^{B}\left(\frac{a_t}{H_{tt}}\right)^2 + \frac{1}{S}\left(\partial_x B - \frac{\partial_xS}{S}\right)^2\right].
\end{align}\end{subequations}
Inverting this matrix returns (\ref{donosdata1}), the exact result found holographically in \cite{donos1409}.     

More recently \cite{rangamani} has generalized the results of \cite{donos1409} to include the effects of a dynamical scalar field in the dual theory.   In general this scalar must be consistently included within hydrodynamics, and so we must consider a more general theory to connect with these results in the generic case.

%

In the case where translational symmetry is broken in multiple directions, there is an important subtlety.  It turns out that the local ``current" in the emergent horizon fluid is \emph{not} equivalent to $\langle J\rangle$ in the boundary theory.   This is true only after taking a spatial average \cite{donos1506}.   To relate the ``current" in the horizon fluid to the current the boundary fluid, one must add a non-local integral over the bulk direction.

It was recently shown \cite{donos1506} in a more general context that dc transport in holography reduces to solving ``hydrodynamic" equations on the black hole horizon.   \cite{donos1506} interpreted the resulting fluid equations as an incompressible Navier-Stokes equation.  \cite{grozdanov} points out that these equations can also be interpreted in the framework of the present paper.

These  examples suggest that -- while the hydrodynamic framework of this paper is extremely helpful providing some physical intuition to these non-perturbative holographic results -- this story is not complete.   Importantly, however, much of the variational technology that we develop can be directly applied to holographic models.  

\section{Strong Disorder}\label{sec5}
We cannot be as rigorous in the strong disorder limit and give closed form expressions for the conductivity matrix.   Nonetheless, we will develop simple but powerful variational methods that allow us to get a flavor of transport at strong disorder, by providing lower and upper bounds on the conductivity matrix.      

We focus on the discussion of $\sigma^{\mathrm{qq}}_{ij}$ in an isotropic theory in this section.   However, the techniques developed below may be used to compute all thermoelectric transport coefficients.   Our discussion is therefore not an exhaustion of all possible physics contained in the hydrodynamic formalism, but simply a demonstration of what we believe is a general feature of hydrodynamic transport:  a crossover from coherent (Drude) physics to incoherent  behavior as disorder strength increases.   

We present the mathematical formalism in the subsections below -- explicit examples of calculations may be found in Appendix \ref{appa}.

\subsection{Power Dissipated}
Define ``voltage drops" $V^\alpha_i$ of each conserved quantity in each direction as \begin{equation}
V^\alpha_i \equiv  \Phi^\alpha (x_i=0, \mathbf{x}_{\perp i}) -  \Phi^\alpha(x_i=L, \mathbf{x}_{\perp i}).
\end{equation}
Recall our boundary conditions on $\Phi^\alpha$: it must be periodic up to linear terms. We also define the net currents flowing in the $i$ direction via \begin{equation}
I^\alpha_i \equiv \int\limits_{\text{fixed }x_i} \mathrm{d}^{d-1}\mathbf{x}\; \mathcal{J}^\alpha_i.
\end{equation}Note that by current conservation, $I^\alpha_i$ can be evaluated at any $x_i$.   Since $V$ and $I$ are determined by the solution to a linear response problem, we can relate them via a conductance matrix (do not confuse this $G^{\alpha\beta}$ with the retarded Green's function defined previously)\begin{equation}
I^\alpha_i \equiv G^{\alpha\beta}_{ij} V^\beta_j,
\end{equation}or via its inverse, the resistance matrix \begin{equation}
V^\alpha_i = R^{\alpha\beta}_{ij} I^\beta_j.
\end{equation}  $G^{\alpha\beta}$ is by definition related to the dc transport coefficients: \begin{equation}
G^{\alpha\beta}_{ij} = L^{d-2} \sigma^{\alpha\beta}_{ij}.
\end{equation}
  We claim that the power dissipated in the system is simply given by \begin{equation}
\mathcal{P} = I^\alpha_i R^{\alpha\beta}_{ij} I^\beta_j = V^\alpha_i G^{\alpha\beta}_{ij} V^\beta_j.
\end{equation}

Let us verify this.  Energy is dissipated\footnote{Of course, this would lead to temperature growth at second order in perturbation theory, so that the energy conservation equation (up to external sources) exactly holds at all orders.} locally via the dissipative ($\Sigma$ and $\eta$) terms in hydrodynamics: \begin{equation}
\mathcal{P} = \int \mathrm{d}^d\mathbf{x} \left(\Sigma_{ij}^{\alpha\beta} \partial_i\Phi^\alpha \partial_j \Phi^\beta + \eta_{ijkl} \partial_j v_i \partial_l v_k\right).   \label{eqppd}
\end{equation} We integrate by parts on the second term and use (\ref{maineq2}) (recall $v_i$ obeys periodic boundary conditions): \begin{equation}
\mathcal{P} = \int \mathrm{d}^d\mathbf{x} \left(\Sigma_{ij}^{\alpha\beta} \partial_i\Phi^\alpha \partial_j \Phi^\beta - v_i \rho^\alpha \partial_i\Phi^\alpha \right) = - \int \mathrm{d}^d\mathbf{x} \; \mathcal{J}^\alpha_i \partial_i\Phi^\alpha.
\end{equation}But $\mathcal{J}^\alpha_i$ is a conserved current, and so \begin{equation}
\mathcal{P} = -\oint \mathrm{d}^{d-1}\mathbf{x} \; \Phi^\alpha n_i \mathcal{J}^\alpha_i = \sum_i I^\alpha_i(\Phi^\alpha(x_i=0) - \Phi^\alpha(x_i=L)) = I^\alpha_i V^\alpha_i.
\end{equation}with $n_i$ the outward pointing normal.

\subsection{Lower Bounds}
Let us begin by discussing the lower bounds on conductivities.   These are by far the more important bounds to obtain, because -- as we will see -- they allow us to rule out insulating behavior in a wide variety of strongly disordered hydrodynamic systems.   

We obtain lower bounds on conductivities analogous to how one obtains upper bounds on the resistance of a disordered resistor network, via Thomson's principle \cite{levin}.   Similar approaches are also used in kinetic theory \cite{ziman}.  Thomson's principle states that if we run any set of ``trial" currents through a resistor network, subject to appropriate boundary conditions, then we can upper bound the inverse conductivity by simply computing the power dissipated by our trial currents.   The power dissipated in the resistor network is minimal on the true distribution of currents, which is compatible with Ohm's Law and a singly-valued voltage function.  We will see that remarkably, this simple approach immediately generalizes.  

Let us propose a trial set of charge and heat currents,  $\tilde{\mathcal{J}}^\alpha_i$, which are periodic functions, and exactly conserved: \begin{equation}
\partial_i \tilde{\mathcal{J}}^\alpha_i =0.
\end{equation}
 In general, this trial function will not be compatible with a single-valued (well-defined) $\Phi^\alpha$.  We write \begin{equation}
\tilde{\mathcal{J}}^\alpha_i = \bar{\mathcal{J}}^\alpha_i  + \hat{\mathcal{J}}^\alpha_i
\end{equation}with overbars denoting the true solution of the hydrodynamic equations subject to our boundary conditions, tildes denoting our trial ``guesses" at the true solution, and hats denoting the deivations, on all variables henceforth.   We also impose \begin{equation}
\int \mathrm{d}^{d-1}\mathbf{x}\;  \hat{\mathcal{J}}^\alpha_i(x_i=0,L) = 0,   \label{hatjaeq}
\end{equation}
as there is a true solution $\bar{\mathcal{J}}^\alpha_i$ with the same net currents $I^\alpha_i$ as our trial $ \tilde{\mathcal{J}}^\alpha_i$.
We also propose a completely arbitrary periodic velocity field $\tilde{v}_i$.   Define \begin{equation}
\tilde{\mathcal{P}} = \int \mathrm{d}^d\mathbf{x}\left[\left(\Sigma^{-1}\right)^{\alpha\beta}_{ij}\left(\tilde{\mathcal{J}}^\alpha_i - \rho^\alpha \tilde{v}_i\right)\left(\tilde{\mathcal{J}}^\beta_j - \rho^\beta \tilde{v}_j\right) + \eta_{ijkl} \partial_j \tilde{v}_i \partial_l \tilde{v}_k \right],  \label{tildeplower}
\end{equation}which, on the true solution, is analogous to (\ref{eqppd}).   We define $\bar{\mathcal{P}}$ (the true power dissipated) and $\hat{\mathcal{P}}$ analogously.   Recall $\bar{\mathcal{P}},\;\tilde{\mathcal{P}},\;\hat{\mathcal{P}}\ge 0$, and expand out $\tilde{\mathcal{P}}$: \begin{equation}
\tilde{\mathcal{P}} = \hat{\mathcal{P}} + \bar{\mathcal{P}} +2 \int \mathrm{d}^{d}\mathbf{x}\; \left[\left(\Sigma^{-1}\right)^{\alpha\beta}_{ij}\left(\hat{\mathcal{J}}^\alpha_i - \rho^\alpha \hat{v}_i\right)\left(\bar{\mathcal{J}}^\beta_j - \rho^\beta \bar{v}_j\right) + \eta_{ijkl} \partial_j \hat{v}_i \partial_l \bar{v}_k \right] \equiv \hat{\mathcal{P}} + \bar{\mathcal{P}} +2\mathcal{K}
\end{equation}Now \begin{equation}
\left(\Sigma^{-1}\right)^{\alpha\beta}_{ij}\left(\hat{\mathcal{J}}^\alpha_i - \rho^\alpha \hat{v}_i\right)\left(\bar{\mathcal{J}}^\beta_j - \rho^\beta \bar{v}_j\right) = -\left(\Sigma^{-1}\right)^{\alpha\beta}_{ij}\left(\hat{\mathcal{J}}^\alpha_i - \rho^\alpha \hat{v}_i\right) \Sigma^{\beta\gamma}_{jk} \partial_k \bar{\Phi}^\gamma = -\partial_i \bar\Phi^\alpha\left(\hat{\mathcal{J}}^\alpha_i - \rho^\alpha \hat{v}_i\right)
\end{equation}and so we obtain, integrating by parts:  \begin{align}
\mathcal{K} &= \int \mathrm{d}^d\mathbf{x} \left[ - \partial_i \bar\Phi^\alpha \hat{\mathcal{J}}^\alpha_i + \rho^\alpha \hat{v}_i \partial_i \bar{\Phi}^\alpha + \eta_{ijkl} \partial_j \hat{v}_i \partial_l \bar{v}_k\right]  \notag \\ &= -\oint \mathrm{d}^{d-1}\mathbf{x} \bar{\Phi}^\alpha n_i \hat{\mathcal{J}}_i^\alpha + \int \mathrm{d}^d\mathbf{x} \left[  \rho^\alpha  \partial_i \bar{\Phi}^\alpha - \partial_j ( \eta_{ijkl}  \partial_l \bar{v}_k)\right]\hat{v}_i  = 0.
\end{align}
The first term vanishes since $\tilde{\mathcal{J}}^\alpha_i$ is periodic, and the constant gradient terms vanish due to (\ref{hatjaeq});  the second by (\ref{maineq2}).    We conclude that $\tilde{\mathcal{P}} \ge \bar{\mathcal{P}}$.   If we define $\tilde{\mathcal{P}} = \tilde{R}^{\alpha\beta}_{ij} I^\alpha_i I^\beta_j$, then we obtain \begin{equation}
I^\alpha_i \tilde{R}^{\alpha\beta}_{ij} I^\beta_j \ge I^\alpha_i R^{\alpha\beta}_{ij} I^\beta_j.
\end{equation}

In particular, we can immediately obtain bounds for all diagonal entries of $R^{\alpha\beta}_{ij}$.  Suppose that we have a large, isotropic disordered metal, in which case we find $R^{\alpha\beta}_{ij} = R^{\alpha\beta}\mdelta_{ij}$ and $\tilde R^{\alpha\beta}_{ij} = \tilde R^{\alpha\beta}\mdelta_{ij}$.   Then we have the generic bounds \begin{subequations}\label{eqlowerbound}\begin{align}
\frac{1}{\tilde R^{\mathrm{qq}}}  &\le \frac{1}{R^{\mathrm{qq}}} = \frac{G^{\mathrm{hh}}G^{\mathrm{qq}}- (G^{\mathrm{hq}})^2}{G^{\mathrm{hh}}} \le   G^{\mathrm{qq}}, \\
\frac{1}{\tilde R^{\mathrm{hh}}}  &\le G^{\mathrm{hh}}.
\end{align}\end{subequations}    
It is also straightforward to convert from $R^{\alpha\beta}_{ij}$ into $(\sigma^{-1})^{\alpha\beta}_{ij}$: \begin{equation}
\mathcal{P} = I^\alpha_i R^{\alpha\beta}_{ij}I^\beta_j = L^{2d-2} \mathbb{E}\left[\mathcal{J}^\alpha_i \right] R^{\alpha\beta}_{ij} \mathbb{E}\left[\mathcal{J}^\beta_j \right] = L^d \mathbb{E}\left[\mathcal{J}^\alpha_i \right] (\sigma^{-1})^{\alpha\beta}_{ij} \mathbb{E}\left[\mathcal{J}^\beta_j \right].
\end{equation}
To obtain bounds on off-diagonal elements is slightly more subtle.   Information can be found about off-diagonal elements by studying linear combinations of various components of $I^\alpha_i$, but it is not as clearcut as (\ref{eqlowerbound}).

\subsection{Upper Bounds}
Obtaining upper bounds on conductivities is in principle more simple, but quite a bit more subtle in practice.    Let us write \begin{equation}
\mathcal{P} = \int \mathrm{d}^d\mathbf{x}\left(\Sigma_{ij}^{\alpha\beta}\partial_i \Phi^\alpha \partial_j \Phi^\beta  + \eta^{-1}_{ijkl} \mathcal{T}_{ij}\mathcal{T}_{kl}\right) \equiv V^\alpha_i G^{\alpha\beta}_{ij} V^\beta_j,  \label{pupperbound}
\end{equation}
where $\mathcal{T}_{ij} = -\eta_{ijkl} \partial_l v_k$ is the viscous stress tensor, which has $d(d+1)/2$ independent components;  $\eta^{-1}$ is a matrix inverse with the first two indices grouped together and the last two indices grouped together (but only in symmetric combinations).   It is possible that $\eta$ may not be invertible, but in this case it is straightforward to regulate the zero eigenvalue with an infinitesimal positive eigenvalue and then take the inverse.\footnote{For example, this may correspond in a conformal fluid to deforming the equation of state with a non-zero bulk viscosity.}   To compute the conductance, we need to demand (as stated previously) that $\Phi^\alpha$ obeys $\Phi^\alpha(x_i=0) - \Phi^\alpha(x_i=L) = V^\alpha_i$.    

We are going to guess a single valued trial function $\tilde{\Phi}^\alpha = \bar\Phi^\alpha + \hat\Phi^\alpha$, with $\bar\Phi^\alpha$ the exact solution as before,  and $\hat\Phi$ a periodic function (recall that $\Phi^\alpha$ should be periodic up to the linear gradient terms).   We will also guess a trial $\tilde{\mathcal{T}}_{ij} = \bar{\mathcal{T}}_{ij} + \hat{\mathcal{T}}_{ij}$, which must be a symmetric tensor, and a periodic function.   We do \emph{not} require that $\tilde{\mathcal{T}}_{ij}$ be expressable in terms of a velocity function,  or that $\partial_i\tilde{\mathcal{J}}^\alpha_i = 0$.   Let us verify the circumstances under which we can nevertheless find $\tilde{\mathcal{P}} \ge \bar{\mathcal{P}}$, as before.   We find that $\tilde{\mathcal{P}} = \bar{\mathcal{P}} + \hat{\mathcal{P}} + 2\mathcal{K}$, with \begin{align}
\mathcal{K}  &= \int \mathrm{d}^d\mathbf{x} \left( \Sigma^{\alpha\beta}_{ij} \partial_i \hat\Phi^\alpha \partial_j \bar\Phi^\beta  + \eta^{-1}_{ijkl} \hat{\mathcal{T}}_{ij} \bar{\mathcal{T}}_{kl}\right) = \int \mathrm{d}^d\mathbf{x} \left( \left(\rho^\alpha \bar{v}_i - \bar{\mathcal{J}}^\alpha_i\right) \partial_i \hat\Phi^\alpha  - \eta^{-1}_{ijkl} \hat{\mathcal{T}}_{ij} \eta_{klmn} \partial_n \bar{v}_m\right) \notag \\
&=  \int \mathrm{d}^d\mathbf{x} \left(\rho^\alpha \partial_i \hat{\Phi}^\alpha + \partial_j \hat{\mathcal{T}}_{ij} \right)\bar{v}_i - \oint \mathrm{d}^{d-1}\mathbf{x} \; n_i \bar{\mathcal{J}}^\alpha_i \hat{\Phi}^\alpha
\end{align}
We have used the periodicity of $\tilde{\mathcal{T}}_{ij}$ to integrate that term in $\mathcal{K}$ by parts.   The first term vanishes if we require that \begin{equation}
\rho^\alpha \partial_i \tilde{\Phi}^\alpha + \partial_j \tilde{\mathcal{T}}_{ij} =0  \label{upcons}
\end{equation}
of all perturbations.   The second term vanishes because $\hat\Phi^\alpha$ and $\bar{\mathcal{J}}^\alpha_i$ are periodic, with $n_i \bar{\mathcal{J}}^\alpha_i$ taking opposite signs on each face.

 And since the integrand in (\ref{pupperbound}) is positive semi-definite, $\hat{\mathcal{P}}\ge 0$.    We conclude that this forms the basis of a variational principle for the computation of $G^{\alpha\beta}_{ij}$.    If we define \begin{equation}
\tilde{\mathcal{P}} = V^\alpha_i \tilde{G}^{\alpha\beta}_{ij} V^\beta_j,
\end{equation} then we obtain bounds on $G^{\alpha\beta}_{ij}$, analogously to $\tilde{G}^{\alpha\beta}_{ij}$.   As before, diagonal elements of $G^{\alpha\beta}_{ij}$ can be straightforwardly upper bounded, and off-diagonal elements require more care.

\subsection{Discussion of Variational Results}
Here we present a summary of the calculations performed in Appendix \ref{appa}.  For simplicity, let $\mathbb{E}[\mathcal{Q}]=\mathcal{Q}_0$ and $\mathrm{Var}[\mathcal{Q}] = \mathbb{E}[\mathcal{Q}^2] - \mathbb{E}[\mathcal{Q}]^2 = u^2$.   The electrical conductivity will, in a disordered isotropic fluid without parametrically large fluctuations in $\Sigma$ or $\eta$, be bounded from above and below by the following schematic bounds: \begin{equation}
\sigma_{\textsc{q}1}(u) + \sigma_{\textsc{q}2}(u) \frac{\mathcal{Q}_0^2}{u^2} \le  \sigma \lesssim \sigma_{\textsc{q}3}(u) + \sigma_{\textsc{q}4}(u) \frac{\mathcal{Q}_0^2}{u^2}  + \frac{\xi^2\mathcal{Q}_0^4}{\eta_1(u) u^2} + \frac{\xi^2 u^2}{\eta_2(u)}+ \frac{\xi^2 \mathcal{Q}_0^2}{\eta_3(u)}   \label{boundseq}
\end{equation}with each $\sigma_{\textsc{q}}$ and $\eta$ factor above related to ``typical" behavior of $\Sigma^{\mathrm{qq}}$ and $\eta_{ijkl}$ respectively.    In particular the upper bounds are quite subtle (see (\ref{eq125})), and so each $\sigma_{\textsc{q}1,2,3,4}$ may have complicated $u$ dependence for $u\gtrsim\mathcal{Q}_0$, and we have written (\ref{boundseq}) as we did to emphasize qualitative behavior, as discussed after (\ref{eq2}).

(\ref{boundseq}) proves that there is a crossover at $u\sim \mathcal{Q}_0$ between an coherent regime when $u \ll \mathcal{Q}_0$ (translational symmetry is weakly broken) and an incoherent regime when $u\gg\mathcal{Q}_0$  (translational symmetry is strongly broken), as depicted in Figure \ref{fig1}.   As discussed in the introduction,  this is the  physics found by mean field holographic models,  and demonstration of this without a mean field treatment of disorder is a primary quantitative result of this paper.

Many of the statements which lead to (\ref{boundseq}) can be made quite rigorously.   The lower bounds on conductivity are derived carefully and will be valid in a wide variety of theories.  The upper bounds which we derive are much more challenging to evaluate analytically when viscosity is not neglected, and so we have made heuristic arguments to understand the qualitative physics that are  non-rigorous and may break down in some cases.       Theories with very large fluctuations in $\Sigma$, $\eta$, or $\rho^\alpha$ could render the upper and lower bounds to be far enough apart (perhaps parametrically so) for (\ref{boundseq}) to not be useful.     Still, we propose that the coherent-to-incoherent crossover described by (\ref{boundseq}) is generic, and will provide some intuition into why this occurs.

In (\ref{eq2}), we ignored viscous effects, and in Figure \ref{fig1}, we depicted $\sigma$ saturating at $\sigma^*$ when $u\gg \mathcal{Q}_0$.   (\ref{boundseq}) generically confirms this picture, with $\sigma_{\textsc{q}1} \lesssim \sigma^* \lesssim \sigma_{\textsc{q}3}$, so long as \begin{equation}
\sigma_{\textsc{q}}\eta \gg \xi^2u^2.   \label{5bound1}
\end{equation} This inequality may or may not be satisfied, and determines whether transport may become sensitive to viscous effects.  For example, in a strongly interacting quantum critical system of dynamical exponent $z$,  we expect $\sigma_{\textsc{q}} \sim T^{(d-2)/z}$, $\eta \sim  \mathcal{S}\sim T^{d/z}$ \cite{kss}, and $\xi\gg T^{-1/z}$.    The requirement that (\ref{5bound1}) is violated is \begin{equation}
u \gtrsim \sqrt{\frac{\sigma_{\textsc{q}}}{T^{(d-2)/z}}} \frac{T^{d/z}}{T^{1/z}\xi} \sim \sqrt{\frac{\sigma_{\textsc{q}}}{T^{(d-2)/z}}} \frac{\mathcal{S}}{T^{1/z}\xi}.    \label{5bound2}
\end{equation}
When $\mu\lesssim T$ (the regime of validity\footnote{When $\mu \gg T$, generally Fermi liquid theory is valid.} of the hydrodynamic approach in a typical quantum critical model), then it is reasonable to expect that $u \lesssim \mathcal{S}$, as most of the entropy will be associated with a background charge neutral plasma, and not with the deformation by a chemical potential.   Rearranging (\ref{5bound2}) we find \begin{equation}
1 \gtrsim \frac{u}{\mathcal{S}} \gtrsim \sqrt{\frac{\sigma_{\textsc{q}}}{T^{(d-2)/z}}} \frac{1}{T^{1/z}\xi}.
\end{equation}
 Recalling that $T^{1/z}\xi \gg 1$, this sequence of equalities is satisfied for disorder on the longest wavelengths.   However, if $u\ll \mathcal{S}$ then it may be possible to have disorder on short enough wavelengths that this sequence of equalities is not satisfied.  It is this regime where (\ref{eq2}) and Figure \ref{fig1} are valid.     A viscous-dominated transport regime is not understood well, and a further understanding of this regime is an important goal for future work.
 
 A second assumption that went into (\ref{eq2}) is that $(\Sigma^{-1})^{\mathrm{qq}}$ is finite.  This is true in the effective horizon fluid of holographic models, but need not be true in other quantum critical models.   Transport in models where $(\Sigma^{-1})^{\mathrm{qq}}$ is infinite will be discussed elsewhere.

Similar bounds can be found for other transport coefficients.  In particular, for bounds on $\bar\kappa$, one must simply replace $\sigma_{\textsc{q}}$ with $T\bar\kappa_{\textsc{q}}$,  $\mathcal{Q}_0$ with $T\mathcal{S}_0=\mathbb{E}[T\mathcal{S}]$, and $u^2$ with $T^2 \mathrm{Var}[\mathcal{S}]$.    It is more likely that thermal transport is sensitive to viscosity in a quantum critical system, as $\bar\kappa_{\textsc{q}} \rightarrow 0$ when $\bar \mu \ll T$ \cite{hkms}.


One of the most important results we find is an exact inequality for an isotropic fluid: \begin{equation}
\sigma^{\alpha\alpha} \ge \frac{1}{\mathbb{E}\left[\left(\Sigma^{-1}\right)^{\alpha\alpha}_{ii}\right]} , \;\;\;\text{(no summation on }\alpha\text{ or } i).  \label{saa}
\end{equation}
This can be interpreted simply as the statement that a uniform charge or heat current could flow through the fluid, with no convective transport, encountering this effective conductivity.    And so as long as a current can flow everywhere locally with a finite conductivity, so can a current flow globally.     (\ref{saa}) is incredibly powerful -- in particular, if $\Sigma^{\mathrm{qq}}$ is strictly positive at all points in space, we have \emph{proven} that the QFT described by this framework is a conductor.    As mentioned above, the lower bounds in (\ref{boundseq}) are derived quite carefully, and essentially follow from generalizations of (\ref{saa}).    Our proof that these fluids are conductors when $\Sigma^{\alpha\beta}_{ij}$ is finite generalize to anisotropic theories, though the bounds become more easily expressed as upper bounds on the matrix $\sigma^{-1}$.

The new approaches advocated in this paper, along with the existing mean-field literature, suggests that the fate of most holographic models at strong disorder -- at fixed temperature $T$, and arbitrarily strong disorder -- is to become an incoherent conductor, and not an interacting quantum glass.   This is a remarkable and highly non-trivial prediction.   In contrast,  in metals described by (fermionic) free  quantum field theories, there is a transition to an insulating phase at some critical disorder strength \cite{anderson}, which is zero in $d\le 2$ \cite{abrahams}.    Generically, interactions do lead to delocalized, conducting phases at weak disorder, with localization and insulating physics arising at stronger disorder strengths, in any $d$ \cite{giamarchi, basko}.   It is possible that this localization transition is not observable in classical holography, which only captures the leading order in $N\rightarrow \infty$, and so has taken the ``coupling strength $\rightarrow\infty$" limit before the ``disorder $\rightarrow\infty$" limit.

We have always referred to these hydrodynamic models as incoherent metals.   Holographic ``insulators" discussed in the literature typically rely on $\Sigma^{\mathrm{qq}}$ scaling as a positive power of $T$, in a homogeneous model.    In \cite{rangamani}, it is likely due to stripes of such decreasing $\Sigma^{\mathrm{qq}}$ arising at low $T$.   More generally, such insulators  arise from the percolation of locally insulating $\Sigma^{\mathrm{qq}}$ regions through the effective horizon fluid.\footnote{As the percolation problem is trivial in $d=1$, the study of holographic insulators may be quite a bit richer in models with translational symmetry broken in multiple spatial dimensions.}   This is not unlike the ``metal-insulator" transition of a classical disordered resistor lattice, associated with percolation of $R=\infty$ resistors across the lattice \cite{kirkpatrick, derrida}.   This is a different mechanism from Anderson localization in typical condensed matter systems, which is related to destructive interference of quasiparticles scattering off of disorder.   Of course, in holographic models, the percolation phenomenon on the horizon could emerge from ``benign" disorder on the boundary, but from the point of view of the emergent horizon fluid the metal-insulator transition is simply a percolation transition.    We emphasize that  our hydrodynamic formalism is still mathematically valid for dc transport in holographic insulators,   due to the remarkable mathematical results of \cite{donos1506, donos1507}.   The physical interpretation of such a fluid is an important question for future work, as emphasized in the previous section.    There is to date no construction of a holographic metal-insulator transition that is unambiguously driven by (non-striped) disorder, and interpreting any such model in terms of hydrodynamic transport may lead to interesting insights.    In simple holographic models, it has recently been shown that such a transition is impossible \cite{grozdanov}, and so more complicated models with bulk scalar fields will be necessary.    

In a non-holographic context, it is less clear whether or not our hydrodynamic formalism will be valid in a quantum system undergoing a metal-insulator transition, as the validity of hydrodynamics rests on the disorder being long wavelength.   The classical ``metal-insulator" transition realized by resistor networks \cite{kirkpatrick, derrida} is a crude example of this phenomenon, but relies on the only hydrodynamic degree of freedom being charge.

Finally, as we are studying a strongly disordered system, it is also worthwhile to think about fluctuations in the transport coefficients between different realizations of the quenched disorder.   As in \cite{lucas1411}, we expect that these fluctuations are suppressed as the size of the sample increases with $L$ as $L^{-d/2}$, with possible deviations when distributions on the random coefficients $\rho$, $\Sigma$ and $\eta$ are heavy tailed.  Such fluctuations are classical, but this is not surprising since the dc response of our QFTs are governed by classical hydrodynamics.  This is analogous to weakly interacting theories at finite temperature \cite{leestone2}.           In contrast, a free quantum field theory has universal conductance fluctuations at $T=0$ \cite{leestone, altshuler, imry}, so it would be interesting to ask if the $T\rightarrow 0$ limit of holographic models (where hydrodynamics can still be a sensible approach \cite{davisoncold}) has anomalous fluctuations in transport coefficients, in disordered models.

\section{Localization}
As we previously mentioned, many free or weakly interacting quantum systems are described by a ``localized" phase where transport is exponentially suppressed at low temperatures \cite{anderson}, at strong disorder.    Naively, one might think that a strong coupling analogue of localization -- with the associated reduction in transport -- would exist at strong disorder.   Indeed, \cite{saremi} provided evidence for a possible connection in a holographic model.   In seeming contrast to this, we have rigorously ruled out any insulating, localized phase in our framework (which includes many such holographic models), so long as the quantum critical conductivity is finite everywhere;  the simple holographic models studied in the literature to date are described by our framework, with finite $\Sigma^{\mathrm{qq}}$ everywhere in space, in most models.

This is consistent with known results in elastic networks and other random resistor networks.  Despite  the localization of classical eigenfunctions \cite{anderson2, john, ludlam}, diffusion and transport are possible even with localized eigenmodes of the linearized hydrodynamic equations.   This has been shown in similar models without convective transport \cite{halperin2, ziman2, amir1, amir2}.   Localization is more subtle in these systems due to the presence of zero modes of the hydrodynamic operators, due to exact conservation laws.   Together with modes of arbitrarily long correlation length with finite eigenvalues, transport is possible despite classical localization, and so the signatures observed in \cite{saremi} need not be important for dc transport.   

The finite momentum or finite frequency response of the system may be more sensitive to localization.  In a simple model of disordered RC circuits, interesting new universal phenomena arise \cite{amir3}.    It is worthwhile to understand finite $\omega$ transport in the class of models in this paper as well.  In particular, it is interesting to ask whether at strong disorder, the Drude peak found via memory matrices \cite{lucasMM} broadens out enough to look ``incoherent" \cite{hartnoll1}, at least at small frequencies, or whether more exotic phenomena emerge.

There is one other point worth making about localized eigenmodes.  In a translationally invariant fluid, long-time tails in hydrodynamic correlation functions in  $d\le 2$   spoil hydrodynamic descriptions of dc transport \cite{kovtunlec}.  In particular, in $d=2$, the conductivity $\sigma(\omega)$ in an uncharged, translationally invariant system picks up a correction $\sim \log (T/|\omega|)$, which diverges as $\omega\rightarrow 0$ \cite{willwk}.   In holography, it is known that such long time tails are quantum bulk effects \cite{caronhuot}, and are thus completely suppressed in the models described in Section \ref{sec4}.    Since  it has been argued that the memory matrix approach gives sensible predictions for realistic strange metal physics \cite{raghu2, patel, debanjan}, and the memory matrix framework employed there can be interpreted hydrodynamically, one might be, a priori, concerned about whether long time tails can spoil dc transport in these models.    If in the thermodynamic limit, all modes (except for the two zero modes) of the classical hydrodynamic operators are localized, as is believed in $d\le 2$ (where long time tails are problematic), then the standard argument for long time  tails \cite{kovtunlec} seems to fail.  It would be interesting to explore this point further in future work.



\section{Conclusion}
In this paper, we have explored the consequences of hydrodynamics on the transport coefficients of a strongly coupled QFT, disordered on large length scales.    We demonstrated that hydrodynamics can be used to understand the memory function computations of momentum relaxation times, which have previously been derived using an abstract and opaque formalism.    It is also straightforward -- at least in principle -- to compute transport coefficients at higher orders in perturbation theory, whereas memory function formulas only give leading order transport coefficients.   Remarkably,  we also demonstrated that  many non-perturbative holographic dc transport computations can be interpreted entirely by solving a hydrodynamic response problem of a new emergent horizon fluid.   Thus, the  technology of Appendix \ref{appa} may be applied to these models, when exact solutions are not available.   We still need specific microscopic theories to compare with $T$-scaling laws in experiments, but this work provides important physical transparency to a large body of recent literature on transport in strongly coupled QFTs.   We emphasize again that the memory function formalism and holography are valid in regimes where hydrodynamics should formally break down, and so it is strange (but useful) that hydrodynamic technology (which is readily understandable) can be used to help interpet these results nonetheless.

The fact that this hydrodynamic framework can be used to interpret such a wide variety of results from memory function or holographic computations is suggestive of the fate of such theories at strong disorder.   Shortly after this paper was released, it was proved in \cite{grozdanov} that all $\mathrm{AdS}_4$-Einstein-Maxwell holographic models are electrical conductors, using these hydrodynamic techniques (which are valid so long as the bulk black hole horizon is connected).   Thus, we do \emph{not} expect a holographic analogue of a many-body localized phase \cite{basko} to exist in many strongly disordered holographic systems.  Strongly disordered black holes have been numerically constructed recently \cite{santosdisorder1, santos, santosdisorder2}; hence, dc transport coefficients in these backgrounds, along with finite momentum or frequency response, may be numerically computable in the near future.      

More generally, we have also demonstrated -- without recourse to mean field treatments of disorder, or to holography -- a framework which generically gives rise to both a coherent metal at weak disorder, and an incoherent metal at strong disorder.   Incoherent metallic physics has been proposed recently to be responsible for some of the exotic thermoelectric properties of the cuprate strange metals \cite{cupratescale1, cupratescale2}.    One can easily imagine taking realistic scaling laws from microscopic models appropriate for cuprates \cite{raghu2, patel, debanjan, patel2} and making quantitative scaling predictions about the strong disorder regime using insights from our framework.    


Hydrodynamics provides a valuable framework for interpreting more specific microscopic calculations.   There are many natural extensions of this work:  two examples are the study of hydrodynamic transport in disordered superfluids and superconductors, or the study of systems perturbed by further deformations than $\bar\mu$.   Still, this framework has limitations.       High frequency transport (in particular, $\omega \gtrsim T$) cannot be captured by hydrodynamics, and provides a unique opportunity for holography in particular to make experimentally relevant predictions about quantum critical dynamics \cite{willwk, prokofev}.

\addcontentsline{toc}{section}{Acknowledgements}
\section*{Acknowledgements}
I would like to thank Ariel Amir, Richard Davison, Blaise Gout\'eraux, Sarang Gopalakrishnan, Bertrand Halperin, Sean Hartnoll and Michael Knap for helpful discussions, and especially Subir Sachdev for  critical discussions on presenting these ideas in a more transparent way.

\begin{appendix}

\section{Onsager Reciprocity}\label{apponsager}
  In this appendix we prove that the thermoelectric conductivity matrix $\sigma^{\alpha\beta}_{ij}$ is symmetric.   This follows entirely from the symmetries of the diffusive transport coefficients (\ref{diffsym1}) and (\ref{diffsym2}), as well as the equations of motion (\ref{maineq1}) and (\ref{maineq2}).   
 
 To do this, we look for a periodic solution $\Phi^\alpha$ and $v_i$,  up to the constant linear terms in $\Phi^\alpha$.   In particular, we write \begin{subequations}\begin{align}
 \Phi^\alpha &=  -F^\alpha_i x_i  + \Phi^{\alpha\beta}_j F^\beta_j, \\
 v_i &=  v^\beta_{ij}F^\beta_j,
 \end{align}\end{subequations}
 which we may always do, as the equations of motion are linear.   (\ref{maineq1}) and (\ref{maineq2}) become:  \begin{subequations}\label{onsagereq}\begin{align}
 \partial_i \left(\rho^\alpha v^\beta_{ij} - \Sigma^{\alpha\gamma}_{ik}\partial_k \Phi^{\gamma\beta}_j\right) &= -\partial_i \Sigma^{\alpha\beta}_{ij}, \\
 \rho^\alpha \partial_i \Phi^{\alpha\beta}_j - \partial_m\left(\eta_{imkl}\partial_l v_{kj}^\beta\right) &= \rho^\beta\mdelta_{ij}.
 \end{align}\end{subequations}
 We further have \begin{equation}
 \sigma^{\alpha\beta}_{ij} = \mathbb{E}\left[\rho^\alpha v^\beta_{ij} - \Sigma^{\alpha\gamma}_{ik}\partial_k \Phi^{\gamma\beta}_j + \Sigma^{\alpha\beta}_{ij}\right] = \mathbb{E}\left[\rho^\alpha v^\beta_{ij} + \Phi^{\gamma\beta}_j  \partial_k\Sigma^{\alpha\gamma}_{ik} + \Sigma^{\alpha\beta}_{ij}\right] .
 \end{equation}
 as we can now always integrate by parts inside of spatial averages.   Now, let us employ (\ref{onsagereq}) and (\ref{diffsym1}) and write \begin{equation}
  \sigma^{\alpha\beta}_{ij} = \mathbb{E}\left[ v^\beta_{kj} \rho^\gamma \partial_k \Phi^{\gamma \alpha}_{ki} + \eta_{klmn}\partial_l v_{kj}^\beta \partial_n v_{mi}^\alpha   +\partial_k\Phi^{\gamma\beta}_j  \left(\rho^\gamma v^\alpha_{ki} - \Sigma^{\gamma\delta}_{kl}\partial_l \Phi^{\delta\alpha}_i\right)+ \Sigma^{\alpha\beta}_{ij} \right].
 \end{equation}
 Using (\ref{diffsym1}) and (\ref{diffsym2}) it is straightforward to see from the previous equation that $\sigma^{\alpha\beta}_{ij}$ is symmetric.
 
 \section{Perturbative Expansions}\label{apppert}
 Let us describe how to extend the weak disorder calculations of Section \ref{sec3} to arbitrarily high orders in perturbation theory, in the special case where the disorder is introduced entirely through $\bar \mu$.   This also gives a flavor for how to ``extend the memory matrix formalism" beyond leading order in perturbation theory.
 
Let us write $\mu = \mu_0 +\epsilon \hat\mu(\mathbf{x})$, with $\epsilon \ll 1$ a perturbatively small number.   Within linear response, the fields $\Phi^\alpha$ and $v_i$ may be written as follows: \begin{subequations}\begin{align}
\Phi^\alpha &= -F^\alpha_ix_i + \sum_{n=-1}^\infty \epsilon^n \Phi^\alpha_{(n)}, \\
v_i &= \sum_{n=-2}^\infty \epsilon^n \bar{v}_{i(n)}  + \sum_{n=-1}^\infty \epsilon^n \tilde{v}_{i(n)}
\end{align}\end{subequations}
where $\mathbb{E}[\tilde{\mathbf{v}}]=\mathbf{0}$,  $\bar{\mathbf{v}}$ a constant, and $\Phi^\alpha_{(n)}$ single-valued.    We will justify the powers of $\epsilon$ above in our computation below, but for now let us emphasize that $\bar v_{i(n-1)}$, $\tilde{v}_{i(n)}$ and $\Phi^\alpha_{(n)}$ enter the computation at the same order.    In addition, the hydrodynamic background becomes disordered: \begin{subequations}\begin{align}
\rho^\alpha &= \rho_0^\alpha + \sum_{n=1}^\infty \epsilon^n \rho_{(n)}^\alpha, \\
\Sigma^{\alpha\beta}_{ij} &= \Sigma^{\alpha\beta}_0 \mdelta_{ij} +  \sum_{n=1}^\infty \epsilon^n \Sigma_{ij(n)}^{\alpha\beta}, \\
\eta_{ijkl} &= \eta_{0ijkl} + \sum_{n=1}^\infty \epsilon^n \eta_{ijkl(n)}.
\end{align}\end{subequations}The background at $\mathrm{O}(\epsilon^0)$ is translation invariant, but not at higher orders.    As in the main text, we will assume isotropy of the leading order transport coefficeints.

(\ref{maineq1}) and (\ref{maineq2}) may be perturbatively expanded in powers of $\epsilon$.  We find the following equations in Fourier space: \begin{subequations}\begin{align}
\mathrm{i}k_i \rho_{(1)}^\alpha(\mathbf{k}) \bar{v}_{i(n-1)} + \mathrm{i}k_i \rho^\alpha_0 \tilde{v}_{i(n)}(\mathbf{k}) + k^2\Sigma_0^{\alpha\beta}\Phi^\beta_{(n)}(\mathbf{k}) &= -X^{\alpha}_{(n)}(\mathbf{k}), \\
\mathrm{i}k_i \rho_0^\alpha \Phi^\alpha_{(n)}(\mathbf{k}) + \eta_0 k_i \left(k_i \tilde{v}_{j(n)}(\mathbf{k}) + k_j \tilde{v}_{i(n)}(\mathbf{k}) - \frac{2}{d}\mdelta_{ij}k_l\tilde{v}_{l(n)}(\mathbf{k})\right) + \zeta_0 \mdelta_{ij}k_l\tilde{v}_{l(n)}(\mathbf{k}) &= -Y_{i(n)}(\mathbf{k}), \\
\mathrm{i}\sum_{\mathbf{k}} k_i \Phi^\alpha_{(n)}(\mathbf{k}) \rho_1^\alpha(-\mathbf{k}) &= -Z_{i(n)}
\end{align}\end{subequations}with the third equation the zero mode of the second, and \begin{subequations}\begin{align}
X^\alpha_{(-1)} &= 0 \\
Y_{i(-1)} &= 0 \\
Z_{i(-1)} &= -\rho_0^\alpha F^\alpha_i
\end{align}\end{subequations}and, for $n\ge 0$:
 \begin{subequations}\begin{align}
X^\alpha_{(n)} &= \mathrm{i}k_i \sum_{m=-2}^{n-2} \rho^\alpha_{(n-m)}(\mathbf{k}) \bar{v}_{i(m)}
+ \mathrm{i}k_i \sum_{m=-1}^{n-1} \rho^\alpha_{(n-m)}(\mathbf{k}-\mathbf{q}) \tilde{v}_{i(m)}(\mathbf{q}) \notag \\
&+ k_i \sum_{m=-1}^{n-1} \sum_{\mathbf{q}} \Sigma_{ij(n-m)}^{\alpha\beta}(\mathbf{k}-\mathbf{q}) q_j \Phi^\beta_{(m)}(\mathbf{q}) \\
Y_{i(n)} &= \sum_{m=-1}^{n-1} \sum_{\mathbf{q}} \left[ \mathrm{i}q_i \Phi^\alpha_{(m)}(\mathbf{q}) \rho^\alpha_{(n-m)}(\mathbf{k}-\mathbf{q}) + \eta_{(n-m)}(\mathbf{k}-\mathbf{q}) k_j q_j \tilde{v}_{i(m)}(\mathbf{q}) \right. \notag \\
&\left. + \left(\eta^\prime_{(n-m)}(\mathbf{k}-\mathbf{q}) - \eta_{(n-m)}(\mathbf{k}-\mathbf{q})\right) k_i q_j \tilde{v}_{j(m)}(\mathbf{q})  \right] \\
Z_{i(n)} &= -\mathrm{i}\sum_{\mathbf{k}} \sum_{m=-1}^{n-1} k_i \Phi^\alpha_{(m)}(\mathbf{k}) \rho^\alpha_{(n+1-m)}(-\mathbf{k}).
\end{align}\end{subequations}
Order by order in perturbation theory, these equations may be solved exactly:  \begin{subequations}\begin{align}
\bar{v}_{i(n-1)} &= -\Gamma^{-1}_{ij} \left(Z_{j(n)} -\mathrm{i} \sum_{\mathbf{k}} k_j \rho^\beta_{(1)}(-\mathbf{k})(\mathfrak{m}(k)^{-1})^{\alpha\beta} \left(X^\alpha_n(\mathbf{k}) - \mathrm{i}\rho_0^\alpha \frac{k_l Y_{l(n)}(\mathbf{k})}{\eta^\prime_0 k^2}\right)\right), \\
\Phi^\alpha_{(n)}(\mathbf{k}) &= -\mathrm{i}(\mathfrak{m}(k)^{-1})^{\alpha\beta}\left(k_i \bar{v}_{i(n-1)} \rho_{(1)}^\beta(\mathbf{k}) + X_{(n)}^\alpha(\mathbf{k}) - \mathrm{i}\rho_0^\beta \frac{k_i Y_{i(n)}(\mathbf{k})}{\eta^\prime_0 k^2}\right), \label{eqphim1} \\
\tilde{v}_{i(n)} &= -\frac{\mathrm{i}}{\eta^\prime_0 k^2} k_i \rho_0^\alpha \Phi^{\alpha}_{(n)}(\mathbf{k})  -\frac{1}{\eta_0 k^2} \left(\mdelta_{ij} - \frac{\eta^\prime_0 - \eta_0}{\eta^\prime_0}\frac{k_ik_j}{k^2}\right) Y_{j(n)}(\mathbf{k})  . 
\end{align}\end{subequations}
with $\Gamma_{ij}$ and $\mathfrak{m}^{\alpha\beta}$ given by (\ref{gammaij32}), with $\hat\rho^\alpha\rightarrow \rho^\alpha_{(1)}$.   These equations have a clearly nested structure and can be iteratively solved.   At leading order, it is readily seen that the response of the fluid is simply \begin{equation}
\mathcal{J}^\alpha_{i(-2)} = \rho_0^\alpha \rho_0^\beta \Gamma^{-1}_{ij} F^\beta_j,
\end{equation}
 as claimed in the main text.   We also stress that even at leading order, $\Phi^\alpha$ and $\tilde{v}_i$ are non-local functions.

At higher orders, the $X$, $Y$ and $Z$ corrections must be systematically accounted for, and this is overwhelming to process by hand, especially without specific equations of state.    However, this tedious procedure does seem easier than attempting to generalize the memory matrix formalism to higher orders in perturbation theory, and indeed makes predictions for such an effort.   So let us at least comment on, qualitatively, what happens at higher orders in perturbation theory.     Many terms that contribute to $\mathcal{J}^\alpha_i$ at higher orders in $\epsilon$ are related to $\rho^\alpha_{(n)}$, $\Sigma^{\alpha\beta}_{ij(n)}$, and $\eta_{(n)}$ and $\zeta_{(n)}$,  at higher powers of $n$.   As we mentioned in Section \ref{sec32}, we can interpret \begin{equation}
\rho^\alpha_{(1)}(\mathbf{k}) = \mathrm{Re}\left[G^{\alpha\mathrm{q}}(\mathbf{k},\omega=0)\right]  \hat\mu(\mathbf{k}) = \chi^{\alpha\mathrm{q}} \hat\mu(\mathbf{k}).
\end{equation}
  Namely, the response coefficients above are related to certain Green's functions that can be computed in a microscopic model.   Recall that the disorder is on such long wavelengths that we may neglect $\mathbf{k}$-dependence in the hydrodynamic Green's functions.  So it is tempting to interpret \begin{align}
\rho^\alpha_{(n)}(\mathbf{k}) &= \frac{1}{n!}\sum_{\mathbf{k}_1,\ldots,\mathbf{k}_{n-1}} \mathrm{Re}\left[ G^{\alpha\mathrm{q}\cdots\mathrm{q}}(\mathbf{k}_1,\ldots,\mathbf{k}_n, \omega=0)\right]  \hat\mu(\mathbf{k}_1)\cdots \hat\mu(\mathbf{k}_n)  \mdelta_{\mathbf{k},\mathbf{k}_1+\cdots+\mathbf{k}_n} \notag \\
&\approx \frac{\chi^{\alpha\mathrm{q}\cdots\mathrm{q}}}{n!} \sum \hat\mu(\mathbf{k}_1)\cdots \hat\mu(\mathbf{k}_n)  \mdelta_{\mathbf{k},\mathbf{k}_1+\cdots+\mathbf{k}_n},
\end{align}
with $G^{\alpha\mathrm{q}}$ an appropriate $n$-point Green's function in the microscopic theory.   Similar statements may be made for $\Sigma$ and $\eta$ by relating them properly to Green's functions of $J_\mu$ and $T_{\mu\nu}$, as in \cite{kovtunlec}.    In the last step, we have used the fact that disorder is long wavelength, and so we expect $\rho^\alpha_{(n)}$, $\Sigma^{\alpha\beta}_{ij(n)}$ and $\eta_{ijkl(n)}$ to be local functions of $\hat\mu$ in position space.

This provides a prediction of our hydrodynamic framework which may be compared with a memory matrix calculation at higher orders in perturbation theory (or another method).   Of course, we should stress that in principle, memory matrix calculations can account for corrections beyond the regime of validity of hydrodynamics,  though in the limits we identified in Section \ref{sec2}, one should find that only the contributions described above contribute to the conductivities.

In the above framework, it does not seem as though there are any natural cancellations between various terms at higher orders in perturbation theory.   So this approach becomes rapidly unwieldy for computing transport coefficients past leading order in $\epsilon$.    Holographic mean field phenomenology suggests that these corrections are all related to a single phenomenological coefficient -- the Drude relaxation time $\tau \sim \epsilon^{-2}$ in (\ref{drude}).   It would be interesting to understand further under what circumstances the Green's functions above undergo similar universal cancellations, and whether this is a sensible prediction of holography.

  \section{Examples of Variational Calculations}\label{appa}

\subsection{Upper Bounds on the Resistance Matrix}
A simple set of trial functions is \begin{subequations}\begin{align}
\tilde{\mathcal{J}}^\alpha_i &= \text{constant}, \\
\tilde{v}_i  &= 0.
\end{align}\end{subequations}
This is a guess corresponding to strong momentum relaxation, as the response of the metal is entirely in the diffusive sector.   Employing (\ref{eqlowerbound}) we obtain (\ref{saa}).

In cases with weak disorder this bound is not strong enough to be useful, and we can do better by allowing for $\tilde{v}_i$ to be a constant ($\mathbf{x}$-independent) variational parameter.   In this case, we obtain \begin{equation}
\tilde{\mathcal{P}}(\tilde{v}_i) = L^d\left[ A^{\alpha\beta}_{ij} \mathcal{J}^\alpha_i \mathcal{J}^\beta_j + 2 B^\beta_{ij} \mathcal{J}^\beta_j \tilde{v}_i + C_{ij}\tilde{v}_i \tilde{v}_j\right]
\end{equation}where \begin{subequations}\begin{align}
A^{\alpha\beta}_{ij} &= \mathbb{E}\left[\left(\Sigma^{-1}\right)^{\alpha\beta}_{ij}\right], \\
B_{ij}^\beta &= -\mathbb{E}\left[\left(\Sigma^{-1}\right)^{\alpha\beta}_{ij} \rho^\alpha\right], \\
C_{ij} &= \mathbb{E}\left[\left(\Sigma^{-1}\right)^{\alpha\beta}_{ij} \rho^\alpha \rho^\beta\right].
\end{align}\end{subequations}
Minimizing $\tilde{\mathcal{P}}(\tilde{v}_i)$, we find \begin{equation}
\mathcal{J}^\alpha_i \left(\sigma^{-1}\right)^{\alpha\beta}_{ij} \mathcal{J}^\beta_j \le \mathcal{J}^\alpha_i\left[ A^{\alpha\beta}_{ij} - B^\alpha_{ik} B^\beta_{jl} C^{-1}_{kl} \right]\mathcal{J}^\beta_j  \equiv  \mathcal{J}^\alpha_i (\tilde{\sigma}^{-1})^{\alpha\beta}_{ij} \mathcal{J}^\beta_j  .   \label{difflower}
\end{equation}

It is straightforward to see that the smallest eigenvalue of $\tilde \sigma^{-1}$ must be larger than the smallest eigenvalue of $\sigma^{-1}$.    A generic consequence of this result is that if the components of $\tilde \sigma^{-1}$ are not parametrically small in the weak disorder limit, the components of $\tilde \sigma$ may be parametrically large in the weak disorder limit.    A simple example analogous to Section \ref{sec32} is the case where $\Sigma$ is a constant, isotropic matrix, and $\rho^\alpha = \rho_0^\alpha + \hat\rho^\alpha$, with $\rho_0^\alpha$ a constant and $\hat\rho^\alpha$ a small perturbation with $\mathbb{E}[\hat\rho^\alpha]=0$.   In this case we find \begin{equation}
(\tilde{\sigma}^{-1})^{\alpha\beta}_{ij} = (\Sigma^{-1})^{\alpha\beta} \mdelta_{ij} - \frac{(\Sigma^{-1})^{\alpha\gamma}(\Sigma^{-1})^{\beta\delta} \rho_0^\gamma \rho_0^\delta }{(\Sigma^{-1})^{\eta\zeta}\left(\rho_0^\eta \rho_0^\zeta + \mathbb{E}\left[\hat\rho^\eta\hat\rho^\zeta\right]\right)}\mdelta_{ij} .
\end{equation}
As $\hat\rho^\alpha \rightarrow 0$, one can check that $\rho_0^\beta$ becomes an eigenvector of $\tilde{\sigma}^{-1}$ with a parametrically small eigenvalue.   Exactly at $\hat\rho^\alpha=0$,  $\sigma^{\alpha\beta}$ will have an eigenvalue of $\infty$; as discussed previously, this follows on quite general principles from the fact that $\mathcal{S}$ and $\mathcal{Q}$ become constant.   If we invert this matrix, we find \begin{equation}
\tilde{\sigma}^{\alpha\beta} \approx  \frac{\rho_0^\alpha \rho_0^\beta}{\mathbb{E}[\hat\rho^\eta \hat\rho^\zeta] (\Sigma^{-1})^{\eta\zeta}} + \cdots  \equiv \frac{\rho_0^\alpha \rho_0^\beta}{\tilde{\mathcal{C}}} + \cdots  \label{sq2}
\end{equation}
To compute this eigenvalue,\footnote{To compute an eigenvalue, one should first properly ``re-dimensionalize" $\hat\rho^\alpha$ so that all matrix elements of $\tilde{\sigma}^{\alpha\beta}$ have the same dimension.} it is easiest to compute $\rho_0^\alpha (\tilde{\sigma}^{-1})^{\alpha\beta}\rho_0^\beta$, and take the leading order coefficient in $\hat\rho^\alpha$.   The subleading corrections correspond to diffusive transport and stay finite in the $\hat\rho^\alpha\rightarrow 0$ limit.

Let us compare to the exact results in the perturbative limit in Section \ref{sec32}.   Technically speaking, we are not guaranteed that $\tilde{\sigma}^{\alpha\beta} \le \sigma^{\alpha\beta}$, though this inequality is satisfied in this limit (assuming that $\rho_0^\alpha>0$).   For we can write $\sigma^{\alpha\beta} \approx \rho_0^\alpha \rho_0^\beta/\mathcal{C}$, with $\mathcal{C}$ given by (\ref{gammaij32}), and  \begin{equation}
\mathcal{C} = \frac{1}{d}\sum_{\mathbf{k}} \hat\rho^\alpha(-\mathbf{k}) \left(\frac{\rho_0^\alpha \rho_0^\beta}{\eta^\prime k^2} + \Sigma^{\alpha\beta}\right)^{-1} \hat\rho^\beta(\mathbf{k}) \le  \frac{1}{d}\sum_{\mathbf{k}}  \hat\rho^\alpha(-\mathbf{k}) (\Sigma^{-1})^{\alpha\beta} \hat\rho^\beta(\mathbf{k}) = \frac{\tilde{\mathcal{C}}}{d}.  
\end{equation}

The first inequality here follows from the fact that for any vector $u_i$, and two positive definite matrices $A_{ij}$ and $B_{ij}$, the following inequality holds: \begin{equation}
u_i (A+B)^{-1}_{ij} u_j \le u_i A^{-1}_{ij}u_j.   \label{matproof1}
\end{equation}To prove this, let $\lambda>0$ be a positive coefficient, and \begin{equation}
\frac{\mathrm{d}}{\mathrm{d}\lambda} u_i (A+\lambda B)^{-1}_{ij} u_j = -u_i (A+\lambda B)^{-1}_{ik} B_{kl} (A+\lambda B)^{-1}_{lj} u_j < 0,  \label{matproof2}
\end{equation}with the latter inequality following from positive-definiteness of sums and products of positive definite matrices.  Integrating (\ref{matproof2}) from $\lambda=0$ to 1 proves (\ref{matproof1}).

It is also possible to find viscosity-limited bounds on the resistance matrix which can be smaller than the diffusion-limited bound (\ref{difflower}), where viscosity plays no role.    For simplicity, let us focus on the specific case of computing thermal transport in an isotropic theory with $\mathcal{Q}=0$.    (\ref{difflower}) gives us that \begin{equation}
(\sigma^{-1})^{\mathrm{hh}} \le \mathbb{E}\left[(\Sigma^{-1})^{\mathrm{hh}}\right] -\frac{\mathbb{E}[(\Sigma^{-1})^{\mathrm{hh}}\mathcal{S}]^2}{\mathbb{E}[(\Sigma^{-1})^{\mathrm{hh}}\mathcal{S}^2]}.   \label{thermb1}
\end{equation}
A natural guess for a viscous-dominated bound is to assume that \begin{subequations}\begin{align}
\tilde{\mathcal{J}}^{\mathrm{h}}_i &= \text{constant}, \\
\tilde{v}_i  &= \frac{\tilde{\mathcal{J}}^{\mathrm{h}}_i}{T\mathcal{S}}.
\end{align}\end{subequations}
This directly leads to \begin{equation}
(\sigma^{-1})^{\mathrm{hh}} \le \mathbb{E}\left[ \frac{\eta^\prime}{d}  \left(\frac{\partial_i\mathcal{S}}{T\mathcal{S}^2}\right)^2 \right].  \label{thermb2}
\end{equation}
We may employ whichever of (\ref{thermb1}) or (\ref{thermb2}) is smaller.

\subsection{Upper Bounds on the Conductivity Matrix}
%

For simplicity, we focus on the bounding of $G^{\mathrm{qq}}_{ij}$;  $G^{\mathrm{hh}}_{ij}$ may be bounded with an exactly analogous ansatz.   Let us write the background charge density as \begin{equation}
\mathcal{Q} = \mathcal{Q}_0 + \hat{\mathcal{Q}},
\end{equation}
with $\mathbb{E}[\mathcal{Q}]=\mathcal{Q}_0 \ne 0$ and $\mathbb{E}[\hat{\mathcal{Q}}]=0$.  Let us also split $\Phi^{\mathrm{q}}$ into a linear term sourcing a background electric field, and a periodic response $\varphi$: \begin{equation}
\Phi^{\mathrm{q}} = \varphi + \sum_i V^{\mathrm{q}}_i \left(1-\frac{x_i}{L}\right).
\end{equation}  It is easier to deal with (\ref{upcons}) in Fourier space, so let us write (with $E_i = V^{\mathrm{q}}_i/L$): \begin{equation}
\mathrm{i}E_i \mathcal{Q}(\mathbf{k}) + \sum_{\mathbf{q}} q_i \varphi(\mathbf{q})\mathcal{Q}(\mathbf{k}-\mathbf{q})  = k_j \mathcal{T}_{ij}(\mathbf{k}).
\end{equation}

Let us begin by assuming that $\eta\rightarrow\infty$, so that we may ignore the response of $\mathcal{T}_{ij}$ to $\Phi$ when computing (\ref{pupperbound}).  
The only constraint we must impose in this limit is \begin{equation}
\mathbb{E}[\mathcal{Q} \partial_i \tilde\Phi^{\mathrm{q}}] = 0.
\end{equation}
A natural guess for $\varphi$, inspired by exact results in the weak disorder limit in Section \ref{sec32}, is \begin{equation}
\varphi(\mathbf{k}) = -\mathrm{i} \frac{k_i E_i }{Ak^2}  \hat{\mathcal{Q}}(\mathbf{k}). \label{eqcalf}
\end{equation}with $A$ a positive constant constrained by (\ref{upcons}): \begin{equation}
E_i \mathcal{Q}_0 = E_j \sum_{\mathbf{q}} \frac{q_iq_j}{Aq^2} |\hat{\mathcal{Q}}(\mathbf{q})|^2  \label{fconstraint}
\end{equation}  
So far, up to neglecting viscous contributions to $\mathcal{P}$, this is completely rigorous.

 For simplicity, suppose that $\hat{\mathcal{Q}}(\mathbf{k})$ are disordered random variables, that are not drawn from a heavy-tailed distribution: \begin{subequations}\begin{align}
 \mathbb{E}_{\mathrm{d}}[\hat{\mathcal{Q}}(\mathbf{k})] &= 0, \\
 \mathbb{E}_{\mathrm{d}}[\hat{\mathcal{Q}}(\mathbf{k})\hat{\mathcal{Q}}(\mathbf{q})] &= \frac{u^2}{N} \mdelta_{\mathbf{k},-\mathbf{q}}.
 \end{align}\end{subequations} This will allow us to extract meaningful qualitative information out of bounds which will be quite opaque in Fourier space.  
$A$ is fixed by (\ref{fconstraint}) as $N\rightarrow\infty$: \begin{equation}
A = \frac{u^2}{d\mathcal{Q}_0}.
\end{equation}
Fluctuations in $A$ are suppressed as $N^{-1/2}$ \cite{lucas1411}.

Plugging this $\Phi$ into (\ref{pupperbound}) we obtain \begin{align}
G^{\mathrm{qq}}_{ij} E_i E_j &\le L^{d-2} \mathbb{E}\left[ \Sigma^{\mathrm{qq}}_{ij}(\mathbf{x}) (E_i - \partial_i \hat \Phi)(E_j - \partial_j \hat \Phi)  \right] \notag \\
&= L^{d-2}\left[ \mathbb{E}\left[\Sigma^{\mathrm{qq}}_{ij}\right] E_i E_j - 2\mathbb{E}\left[\Sigma^{\mathrm{qq}}_{ij}\partial_i \varphi \right] E_j + \mathbb{E}\left[\Sigma^{\mathrm{qq}}_{ij}\partial_i \varphi \partial_j \varphi \right]\right] .
\end{align}
Now, recall that we expect $\Sigma^{\mathrm{qq}}$ is a function of $\bar \mu$, and $\hat{\mathcal{Q}}$ is a function of $\bar \mu$ as well,   so let us just consider $\Sigma^{\mathrm{qq}}$ to be a function of $\hat{\mathcal{Q}}$.   Then we obtain (exploiting isotropy): \begin{align}
\mathbb{E}_{\mathrm{d}}[\sigma^{\mathrm{qq}}_{ij}] &\le \frac{\mdelta_{ij}}{d} \mathbb{E}_{\mathrm{d}} \left[d\Sigma^{\mathrm{qq}}(\hat{\mathcal{Q}}) - 2\sum_{\mathbf{k}}\Sigma^{\mathrm{qq}}(-\mathbf{k})\frac{\hat{\mathcal{Q}}(\mathbf{k})}{A} + \sum_{\mathbf{k}_1,\mathbf{k}_2}\Sigma^{\mathrm{qq}}(-\mathbf{k}_1-\mathbf{k}_2)\frac{\hat{\mathcal{Q}}(\mathbf{k}_1)\hat{\mathcal{Q}}(\mathbf{k}_2)}{A^2}\frac{(\mathbf{k}_1\cdot\mathbf{k}_2)^2}{k_1^2k_2^2}\right] \notag \\
&\le \frac{\mdelta_{ij}}{d}\mathbb{E}_{\mathrm{d}} \left[d\Sigma^{\mathrm{qq}}(\hat{\mathcal{Q}}) - 2\sum_{\mathbf{k}}\Sigma^{\mathrm{qq}}(-\mathbf{k})\frac{\hat{\mathcal{Q}}(\mathbf{k})}{A} + \sum_{\mathbf{k}_1,\mathbf{k}_2}\Sigma^{\mathrm{qq}}(-\mathbf{k}_1-\mathbf{k}_2)\frac{\hat{\mathcal{Q}}(\mathbf{k}_1)\hat{\mathcal{Q}}(\mathbf{k}_2)}{A^2}\right] \notag \\
&= \mathbb{E}_{\mathrm{d}}\left[\Sigma^{\mathrm{qq}} - \frac{2\mathcal{Q}_0\hat{\mathcal{Q}}\Sigma^{\mathrm{qq}}}{u^2} + \frac{d\mathcal{Q}_0^2\hat{\mathcal{Q}}^2 \Sigma^{\mathrm{qq}}}{u^4}\right]\mdelta_{ij}   \label{eq125}
\end{align}
This equation holds for arbitrary functions $\Sigma^{\mathrm{qq}}(\hat{\mathcal{Q}})>0$, so long as viscosity is negligible.   Our disorder-averaged bound on $\sigma_{ij}$ is also manifestly positive, as it must be.

In a theory with $\mathcal{Q}_0 \rightarrow 0$, and viscosity still negligible,  our upper bound collapses to $\mathbb{E}[\Sigma^{\mathrm{qq}}]$.   This bound may also be found at $\mathcal{Q}_0=0$ by directly plugging in the satisfactory ansatz $\Phi^{\mathrm{q}} = -E_ix_i$ into (\ref{pupperbound}), and so our bound is actually valid for all $\mathcal{Q}_0$.

Now, let us consider the effects of finite viscosity.  Henceforth, the discussion will be more qualitative, and we will not be particularly concerned with O(1) prefactors, as it turns out to be quite difficult to write down a good non-perturbative analytic solution to (\ref{upcons}).

Let us see what happens if we simply use (\ref{eqcalf}), along with a sensible ansatz for $\mathcal{T}_{ij}$.   Denoting \begin{equation}
\mathrm{i}E_i \mathcal{Q}(\mathbf{k}) + \sum_{\mathbf{q}}q_i \varphi(\mathbf{q}) \mathcal{Q}(\mathbf{k}-\mathbf{q}) = \mathcal{A}_i(\mathbf{k}),
\end{equation}
we pick in $d=1$: \begin{equation}
\mathcal{T}_{xx} = \frac{\mathcal{A}_x}{k_x},
\end{equation}in $d=2$: \begin{subequations}\begin{align}
\mathcal{T}_{xx} &= - \mathcal{T}_{yy}  = \frac{k_x \mathcal{A}_x - k_y\mathcal{A}_y}{k_x^2+k_y^2}, \\
\mathcal{T}_{xy} &= \frac{k_y\mathcal{A}_x + k_x\mathcal{A}_y}{k_x^2+k_y^2},
\end{align}\end{subequations}and in $d=3$: 
\begin{subequations}\begin{align}
\mathcal{T}_{xy} &= \frac{k_x\mathcal{A}_{x}+k_y\mathcal{A}_{y}-k_z\mathcal{A}_{z}}{2k_xk_y}, \\
\mathcal{T}_{xz} &= \frac{k_x\mathcal{A}_{x}-k_y\mathcal{A}_{y}+k_z\mathcal{A}_{z}}{2k_xk_z}, \\
\mathcal{T}_{yz} &= \frac{-k_x\mathcal{A}_{x}+k_y\mathcal{A}_{y}+k_z\mathcal{A}_{z}}{2k_yk_z}.
\end{align}\end{subequations}
In all of the above cases, the equations are only valid at $\mathbf{k}=\mathbf{0}$, and we take the zero modes to vanish.   In $d=3$, we make the stronger assumption that, e.g., $\mathcal{T}_{xy}=0$ whenever $k_x=0$ or $k_y=0$.   It is now straightforward to (qualitatively) see what happens.   The first contribution to the conductivity is unchaged from (\ref{eq125}),  and the average viscous power dissipated scales as \begin{align}
\mathbb{E}\left[\eta^{-1}_{ijkl}\mathcal{T}_{ij}\mathcal{T}_{kl}\right] &\sim \sum_{\mathbf{k}\ne \mathbf{0}} \eta^{-1}(\mathbf{0}) \frac{\mathcal{A}(\mathbf{k})\mathcal{A}(-\mathbf{k})}{k^2} + \sum_{\mathbf{k}_1,\mathbf{k}_2\ne \mathbf{0}} \eta^{-1}(-\mathbf{k}_1-\mathbf{k}_2)\frac{\mathcal{A}(\mathbf{k}_1)\mathcal{A}(\mathbf{k}_2)}{k_1k_2} \notag \\
&\sim \left[\frac{\xi^2 u^2}{\eta} + \frac{\xi^2\mathcal{Q}_0^2}{\eta} + \frac{\xi^2 \mathcal{Q}_0^4}{\eta u^2}  \right]E_i E_i  \label{eq130}
\end{align}
where we have been schematic and neglected tensor indices on $\eta$.     To obtain the final scaling law above, we have used that $\mathcal{A}(\mathbf{k}\ne \mathbf{0}) \sim \eta^{-1}(\mathbf{k}\ne \mathbf{0}) \sim 1/\sqrt{N}$.   We have neglected in the above scaling the possibility that fluctuations in $\eta$ may be large.   

Of course, the general framework can certainly account for this possibility, and one can directly plug in our ansatzes into (\ref{pupperbound}) -- however, we do not see general simplifcations that can be made, other than the crude scaling arguments here.   Given our ansatzes, the expression in (\ref{eq130}) is generally nonlocal and thus will not be as elegant as (\ref{eq125}).   Putting all of this together, we find (\ref{boundseq}).


Essential to the scaling laws in (\ref{eq130}) was that $\sum k^{-2} \sim N\xi^2$ above.   However, this is not true in $d\le 2$,  where the sum will diverge at small $k$.   We now give a heuristic argument that the typical scaling behavior we found above need not be parametrically different in these dimensions, consistent with what we found previously (in Section \ref{sec32}, the form of the conductivities is the same in all dimensions at weak disorder).   To do so, we need to argue that there is a small modification that we can make to $\varphi$ (so the contribution to the bound on conductivity from $\Sigma^{\mathrm{qq}}$ is qualitatively unchanged),  yet which can remove the IR divergence from the viscous contribution.   The natural guess is to modify $\varphi(\mathbf{k})$ so that $\mathcal{T}_{ij}(\mathbf{k}) =0 $ for $|\mathbf{k}| \xi \lesssim \delta$,   with $\delta \ll 1$ a small constant.   In this case, then we predict that \begin{equation}
\sum \frac{1}{k^2} \sim \left\lbrace\begin{array}{ll}  N/\delta &\ d=1 \\ N\log(1/\delta) &\ d=2\end{array}\right.,
\end{equation}
and so if we choose a ``reasonable" $\delta$  (especially in $d=2$), we can get acceptably finite viscous contributions to the conductivity.    This can be accomplished by writing down \begin{equation}
\varphi = -\mathrm{i}\frac{k_iE_i}{Ak^2}\left(\hat{\mathcal{Q}}(\mathbf{k}) + \delta \cdot Q(\mathbf{k})\right),
\end{equation}with $\varphi_0$ given by (\ref{eqcalf}), and $q\sim \hat{\mathcal{Q}}$,  carrying no anomalous powers of $N$ or $\delta$.   As typical elements of $\varphi$ changed by an amount $\sim \delta$,   we expect that the conductivity in (\ref{eq125}) has only changed by a small amount.

Let us verify this is possible.  We wish to find a solution to the highly overdetermined equation \begin{equation}
 \mdelta_{ij} \hat{\mathcal{Q}}(\mathbf{k}) (1-\mdelta_{\mathbf{k},\mathbf{0}}) - \mathcal{Q}_0 \frac{q_iq_j}{Aq^2}\hat{\mathcal{Q}}(\mathbf{k}) -  \sum_{\mathbf{q}} \hat{\mathcal{Q}}(\mathbf{k}-\mathbf{q}) \frac{q_iq_j}{Aq^2} \hat{\mathcal{Q}}(\mathbf{q}) =\sum_{\mathbf{q}} \mathcal{Q}(\mathbf{k}-\mathbf{q}) \frac{q_iq_j}{Aq^2} \delta Q(\mathbf{q}),\;\;\;\; 0\le |\mathbf{k}| \le \frac{\delta}{\xi}.  \label{eq128}
\end{equation}
The left hand side of the above equation scales as $1/\sqrt{N}$ for typical disorder in $\hat{\mathcal{Q}}$.   In the third term, this scaling follows from the lack of correlations between $\hat{\mathcal{Q}}$ at different momenta.    So long as we choose $Q(\mathbf{k})=0$ in for $|\mathbf{k}|\xi \lesssim 2\delta$, then the ``matrix elements"  $\hat{\mathcal{Q}}$ on the right hand side are also small.   

We now look for $Q(\mathbf{q})$ by solving a constrained optimization problem of the schematic form: \begin{equation}
\mathbf{a} = \mathsf{B}\mathbf{x},
\end{equation} where $\mathbf{a} \in \mathbb{R}^{n_1}$ and $\mathbf{x} \in \mathbb{R}^{n_2}$, $n_1 \sim \delta N \ll n_2\sim (1-2^d\delta)N\sim N$, and $\mathsf{B}$ is a rectangular matrix,  such that $|\mathbf{x}|$ is smallest.   $\mathbf{a}$ is analogous to the (known) left hand side of (\ref{eq128}),   $\mathsf{B}$ is a known, truncated convolution-like matrix in a Fourier basis, and $\mathbf{x}$ plays the role of the undetermined, high momentum modes of $Q(\mathbf{q})\cdot \delta $.   This is a classic problem in constrained optimization with the solution \cite{boyd} \begin{equation}
\delta Q(\mathbf{q}) = \mathbf{a} \cdot (\mathsf{BB}^{\mathrm{T}})^{-1} \mathsf{B}.
\end{equation}
Given that elements of $\mathbf{a}$ and $\mathsf{B}$ each scale as $\sqrt{1/N}$, we roughly estimate that $\delta Q(\mathbf{q}) \sim \delta$, and so indeed it is possible to pick a small correction to $\varphi$ which eliminates the divergence in the viscous contribution to the conductivity.   Note that the matrix $\mathsf{BB}^{\mathrm{T}}$ is nearly diagonal (the off diagonal elements involve uncorrelated sums of random variables and so scale as $1/\sqrt{N}$ instead of 1), and so there are no concerns about parametrically small eigenvalues of $\mathsf{BB}^{\mathrm{T}}$.

 \section{Striped Models}\label{appstripe}
 Thinking about resistivities turns out to be most convenient for models in $d=1$, or with translational symmetry only broken in a single direction $x$, as noted in \cite{andreev}.   This follows from the general arguments that we make in Appendix \ref{appa}.   Since we know that the current flow  $\mathcal{J}^\alpha_x$ is a constant, to solve the variational problem exactly we need only vary (\ref{tildeplower}) with respect to arbitrary $\tilde{v}$ -- the global minimum will corresponds to the true velocity field $\bar v$.   We find that $\bar v$ obeys the differential equation \begin{equation}
\partial_x \left(\eta_{xxxx}\partial_x \bar v\right) = \left(\Sigma^{-1}\right)_{xx}^{\alpha\beta}\left(\mathcal{J}^\beta_x - \rho^\beta \bar v\right)\rho^\alpha.
\end{equation}
This second order linear differential equation cannot be solved exactly in general.  We do not believe that closed form solutions exist in general for $\sigma^{\alpha\beta}_{xx}$, though they can be found in special cases -- for example, if $\Sigma^{\mathrm{hh}}$ is the only non-vanishing diffusive transport coefficient (as in \cite{andreev}, for a non-critical fluid), or if $\Sigma^{\mathrm{qq}}$ is the only non-vanishing coefficient, as in Section \ref{sec:holostripe}.   In both of these cases, $\Sigma^{\alpha\beta}$ is not an invertible matrix -- the zero eigenvector then provides a constraint which fixes $v_x$ in terms of $\mathcal{J}^\alpha_x$.

In particular, let us carry out this computation explicitly for the holographic striped models with equations of state given in Section \ref{sec:holostripe}.   We must generalize the discussion to curved spaces, but this is not so difficult.  The heat conservation equation implies (on a curved space) that \begin{equation}
\mathcal{J}^{\mathrm{h}} = \sqrt{\gamma}\gamma^{xx} T\mathcal{S} v_x = \mathrm{e}^{-B}T\mathcal{S}v_x = \text{constant}.
\end{equation}
Note that $\sqrt{\gamma}=1$, which simplifies calculations.  After an appropriate generalization to curved space, we use (\ref{tildeplower}) (on the true solution) to compute the inverse thermoelectric conductivity matrix:  \begin{align}
\left(\sigma^{-1}\right)^{\alpha\beta}\mathcal{J}^\alpha\mathcal{J}^\beta &= \mathbb{E}\left[\sqrt{\gamma}\left(\gamma_{xx}\left( \mathcal{J}^{\mathrm{q}} - \mathcal{Q} v_x\gamma^{xx}\sqrt{\gamma}\right)^2 + \frac{\eta}{2}\gamma^{ij}\gamma^{kl}s_{ik}s_{jl}\right)\right] \end{align}
where \begin{equation}
s^{ij} \equiv \gamma^{ik}\gamma^{jl} \left(\nabla_k v_l + \nabla_l v_k\right) - \gamma^{ij}\gamma^{kl}\nabla_k v_l.
\end{equation}
For our set-up, parity symmetry in the $y$ direction ensures that $v_y=0$, and that the only non-vanishing components of $s^{ij}$ are \begin{equation}
\gamma_{xx} s^{xx} = -\gamma_{yy}s^{yy} = \mathrm{e}^{-B}\partial_x v_x.
\end{equation}
Putting this together and using (\ref{donosdata2}) to determine $\eta$, $\mathcal{S}$ and $\mathcal{Q}$, we obtain \begin{equation}
\left(\sigma^{-1}\right)^{\alpha\beta}_{xx}\mathcal{J}^\alpha\mathcal{J}^\beta = \mathbb{E}\left[\mathrm{e}^B \left(\mathcal{J}^{\mathrm{q}} - \frac{Sa_t}{4\mpi TSH_{tt}}\mathcal{J}^{\mathrm{h}}\right)^2 + S \left(\mathrm{e}^{-B}\partial_x \left(\frac{\mathrm{e}^B}{4\mpi TS}\right)\right)^2 \left(\mathcal{J}^{\mathrm{h}}\right)^2\right],
\end{equation}which gives (\ref{donosresult}).

  \end{appendix}
\bibliographystyle{unsrt}
\addcontentsline{toc}{section}{References}
\bibliography{disorderbib}

\end{document}